\documentclass[aps,pre,preprint,groupedaddress,showpacs]{revtex4-1}
\usepackage{amssymb}

\usepackage[dvips]{graphicx}
\usepackage{textcomp}
\usepackage{amssymb}

\newcommand{\mc}{\multicolumn}

\begin{document}

\title{
\Large\bf Thermodynamic Casimir effect for films in \\
the three-dimensional Ising  Universality Class: 
Symmetry breaking boundary conditions}

\author{Martin Hasenbusch}
\email[]{Martin.Hasenbusch@physik.hu-berlin.de}
\affiliation{
Institut f\"ur Physik, Humboldt-Universit\"at zu Berlin,
Newtonstr. 15, 12489 Berlin, Germany}

\date{\today}

\begin{abstract}
We study the thermodynamic Casimir force for films in the three-dimensional
Ising universality class with symmetry breaking boundary conditions. 
To this end we simulate the improved Blume-Capel model on the simple 
cubic lattice. We study the two cases $++$, where all spins at the 
boundary are fixed to $+1$ and  $+-$, where the spins at one boundary 
are fixed to $+1$ while those at the other boundary are fixed to $-1$. 
An important issue in analyzing Monte Carlo and experimental data are
corrections to scaling. Since we simulate an improved model, leading 
corrections to scaling, which are proportional to $L_0^{-\omega}$,  
where $L_0$ is the thickness of the film and $\omega \approx 0.8$, can 
be ignored. This allows us to focus on  corrections to scaling  
that are caused by the boundary conditions. The analysis of our data 
shows that these corrections can be accounted for by an effective thickness
$L_{0,eff} = L_0 + L_s$. Studying the correlation length
of the films, the energy per area, the magnetization profile and 
the thermodynamic Casimir force at the 
bulk critical point we find $L_s=1.9(1)$ for our model and the 
boundary conditions discussed here. Using this result for $L_s$ we find
a nice collapse of the finite size scaling curves obtained for the
thicknesses $L_0=8.5$, $16.5$ and $32.5$ for the full range of temperatures 
that we consider.
We compare our results for the finite size
scaling functions $\theta_{++}$ and  $\theta_{+-}$ of the thermodynamic
Casimir force with those obtained in a previous Monte Carlo study,
by the de Gennes-Fisher local-functional method, field theoretic methods
and an experiment with a classical binary liquid mixture.
\end{abstract}
\maketitle

\section{Introduction}
In the thermodynamic limit,
in the neighborhood of a second order phase transition the correlation 
length $\xi$ that is the characteristic length  of thermal fluctuations 
diverges following a power law
\begin{equation}
\label{xipower}
\xi = \xi_{0,\pm} |t|^{-\nu} \times (1 + b_{\pm} |t|^{\theta} + c t + ...)
\;\;,
\end{equation}
where $t=(T-T_c)/T_c$ is the reduced temperature and $\xi_{0,\pm}$ is the 
amplitude of the correlation length in the low ($-$) and the high ($+$) 
temperature phase, respectively. Using this notation, we assume that the high 
temperature phase is characterized by disorder and the low temperature one by 
order. The power law~(\ref{xipower}) is subject to
confluent corrections, such as $b_{\pm} |t|^{\theta}$, and non-confluent ones
such as $c t$. Critical exponents like $\nu$ and ratios of amplitudes such as 
$\xi_{0,+}/\xi_{0,-}$ are universal. This means that they assume exactly the
same value for any system within a given universality class. Also correction 
exponents like $\theta=\omega \nu$ and ratios of correction
amplitudes as $b_+/b_-$ are universal. For the three-dimensional Ising 
universality, which is considered here and other three-dimensional
universality classes like the XY or the Heisenberg universality class, 
$\theta \approx 0.5$. 
%A universality class is characterized by the dimension of the system, the range
%of the interaction and the symmetry of the order parameter. 
For reviews on critical phenomena and the Renormalization Group (RG) see e.g.
\cite{WiKo,Fisher74,Fisher98,PeVi02}.

In 1978  Fisher and de Gennes \cite{FiGe78} realized that when thermal
fluctuations are restricted by a container, a force acts on its walls.
Since this effect is  analogous to the Casimir effect,
where the restriction of quantum fluctuations induces a force, it is called
``thermodynamic'' Casimir effect. Since thermal fluctuations only extend to
large scales in the neighborhood of continuous phase transitions it is
also called ``critical'' Casimir effect. Recently this force could be
detected for various experimental systems and quantitative predictions could
be obtained from Monte Carlo simulations of spin models \cite{Ga09}.

Here we study the thermodynamic Casimir force for the  film geometry.
From a thermodynamic point of view, the thermodynamic Casimir force per area 
is given by 
\begin{equation}
\label{defineFF}
 F_{Casimir} = -   \frac{ \partial \tilde f_{ex} }{ \partial L_0} \;\;,
\end{equation}
where $L_0$ is the thickness of the film and 
$\tilde f_{ex} = \tilde f_{film} - L_0 \tilde f_{bulk}$
is the excess free energy per area of the film, where $\tilde f_{film}$ is
the free energy per area of the film and $\tilde f_{bulk}$ the free energy 
density of the bulk system.
The thermodynamic Casimir force  per area follows the finite size scaling law
\begin{equation}
\label{FCffs}
 F_{Casimir} \simeq k_B T L_0^{-3} \; \theta(t [L_0/\xi_{0,+}]^{1/\nu}) \;\;,
\end{equation}
see e.g. ref. \cite{Krech}. 
The finite size scaling function  $\theta(x)$ depends on the universality
class of the bulk phase transition, the geometry of the finite system and 
the surface universality classes of the 
boundary conditions that are applied.  For reviews of surface 
critical phenomena see \cite{BinderS,Diehl86,Diehl97}.
Similar to the power law~(\ref{xipower}), finite size scaling equations such 
as eq.~(\ref{FCffs}) are subject to corrections to scaling.
In the generic case one expects that leading corrections are  
$\propto L_0^{-\omega}$ (Ref. \cite{Barber}), where $\omega = 0.832(6)$ 
(Ref. \cite{mycritical}) for the three-dimensional Ising universality class. 
Furthermore one expects corrections that are caused by the boundaries. 
We shall give a more detailed discussion of corrections 
to scaling below in section \ref{FFSsection}.

Here we compute finite size scaling functions $\theta$ of the thermodynamic
Casimir force
for the three-dimensional Ising universality class and symmetry breaking 
boundary conditions. Experimentally this situation is realized for example
by a film of a classical binary liquid mixture. Typically, the surface is more 
attractive for one of the two components of the mixture, breaking the symmetry
at the boundary. In the Ising model this can be described by an external field 
that acts on the spins at the surface of the lattice.
Following the classification of surface critical phenomena
such surfaces belong to the normal surface universality
class, which is equivalent to the extraordinary surface universality class
\cite{BuDi94}. In recent experiments on colloidal particles immersed 
in a binary mixture of fluids \cite{NeHeBe09},
the authors have demonstrated that the adsorption
strength can be varied continuously by a chemical modification of the 
surfaces. In particular the situation of effectively equal adsorption 
strengths for the two fluids can be reached. For sufficiently small
ordering interaction at the surface, this corresponds to the ordinary
surface universality class.  Hence these experiments open the way to 
study the crossover between different surface universality classes.
For a recent theoretical discussion of the crossover behaviors of the  
thermodynamic Casimir force see \cite{MoMaDi10} and references therein.
Here we shall not study such crossover behaviors and restrict ourself
to compute finite size scaling functions for the  normal or 
extraordinary universality class. Note that the breaking of the 
effective symmetry between the components of the fluid, or the breaking 
of the $Z_2$ symmetry between $+$ and $-$ spins at the surface in the
Ising model, constitutes a relevant perturbation at the ordinary
fixed point \cite{BinderS,Diehl86,Diehl97}.
Therefore, even for a small breaking of the symmetry, for sufficiently 
large distances, which means in our context a large thickness of the film,
the physics in the neighborhood of the critical point  
is governed by the normal or extraordinary universality class.

Since a film has two surfaces, we can distinguish the two principal
cases: Firstly both boundaries  
attract positive spins, denoted by $++$ in the following, and
secondly one boundary attracts positive spins, while the other 
attracts negative spins, denoted by  $+-$ in the following.
Note that by symmetry $--$ and $-+$ boundary conditions
are equivalent to $++$ and $+-$ boundary conditions,
respectively.

In previous Monte Carlo studies \cite{VaGaMaDi07,VaGaMaDi08}  
the spin-1/2 Ising model has been simulated. Computing finite size scaling 
functions from numerical data obtained for finite thicknesses $L_0$, 
corrections to scaling are a major obstacle. The results for $\theta_{++}$ and
$\theta_{+-}$ given by \cite{VaGaMaDi07,VaGaMaDi08} depend quite strongly on 
the ansatz that is chosen for the corrections.
Here we shall study the improved Blume-Capel model on the simple cubic 
lattice. The  Blume-Capel model is a generalization of the Ising model. 
In addition to $\pm 1$, as in the Ising model, the spin might assume the 
value $0$. The parameter $D$ of the model controls the relative weight of
$0$ and $\pm 1$. For a precise definition see section \ref{themodel} below.
Improved means that the amplitude of corrections $\propto L_0^{-\omega}$
vanishes or in practice is very small compared with the spin-1/2
Ising model.  
Studying thin films this is a quite useful property, since the boundary 
conditions cause corrections that are $\propto L_0^{-1}$ as we shall discuss
below.  Fitting numerical data, it is quite difficult to disentangle
corrections that have similar exponents.
Avoiding this problem we are able to compute the finite size scaling 
functions  $\theta_{++}$ and $\theta_{+-}$ with a small and, as we 
shall argue, reliable error estimate.
Reliable numerical calculations are important, since field theoretic methods
do not provide quantitatively accurate results for the scaling functions 
$\theta_{++}$ and $\theta_{+-}$
as we shall see below. Recently the scaling function $\theta_{++}$ 
has been computed by using the de Gennes-Fisher local-functional method
\cite{BoUp08}. We find a rather good agreement with our result.

The outline of the paper is the following.  First we define the model and
the observables that we have studied. Then we discuss finite size 
scaling and corrections to finite size
scaling. Next we exploit the relation of the spectrum of the transfer matrix 
and the thermodynamic Casimir force. 
Then we discuss the Monte Carlo algorithms that we have used.
We analyze our data obtained from simulations at the critical point 
of the bulk system. This way
we obtain accurate results for the Casimir amplitudes and for $L_s$ that
characterizes the corrections to scaling caused by the boundary conditions.
Next we have simulated in a large range of temperatures around the 
bulk critical point. Based on these simulations we obtain the finite 
size scaling functions $\theta_{++}$ and $\theta_{+-}$ of the thermodynamic
Casimir force. In addition we compute the finite size scaling functions 
of the correlation length of the films. Finally we compare our results
with those obtained by field theoretic methods, the local-functional method,
previous Monte Carlo studies of the Ising model and an experiment on 
a classical binary liquid mixture.

\section{Model}
\label{themodel}
We study the Blume-Capel model on the simple cubic lattice. 
It is defined by the reduced Hamiltonian 
\begin{equation}
\label{Isingaction}
H = -\beta \sum_{<xy>}  s_x s_y
  + D \sum_x s_x^2  \;\; ,
\end{equation}
where the spin might assume the values $s_x \in \{-1, 0, 1 \}$. 
$x=(x_0,x_1,x_2)$
denotes a site on the simple cubic lattice, where 
$x_i \in \{1,2,...,L_i\}$  
and $<xy>$ denotes a pair of nearest neighbors on the lattice.
The inverse temperature is denoted by $\beta=1/k_B T$. The partition function 
is given by $Z = \sum_{\{s\}} \exp(- H)$, where the sum runs over all spin
configurations. The parameter $D$ controls the
density of vacancies $s_x=0$. In the limit $D \rightarrow - \infty$
vacancies are completely suppressed and hence the spin-1/2 Ising
model is recovered.

In  $d > 1$  dimensions the model undergoes a continuous phase transition
for $-\infty \le  D   < D_{tri} $ at a $\beta_c$ that depends on $D$.
For $D > D_{tri}$ the model undergoes a first order phase transition.
The authors of \cite{DeBl04} give for the three-dimensional simple cubic 
lattice $D_{tri}=2.0313(4)$.

Numerically, using Monte Carlo simulations it has been shown that there 
is a point $(D^*,\beta_c(D^*))$ 
on the line of second order phase transitions, where the amplitude
of leading corrections to scaling vanishes.  Our recent 
estimate is $D^*=0.656(20)$ (Ref. \cite{mycritical}). In \cite{mycritical} we
have simulated the model at $D=0.655$ close to $\beta_c$ on lattices of a 
linear size up to $L=360$. From a standard finite size scaling analysis 
of phenomenological couplings like the Binder cumulant we find 
$\beta_c(0.655)=0.387721735(25)$. Furthermore the amplitude of leading 
corrections to scaling is at least by a factor of $30$ smaller than 
for the spin-1/2 Ising model. 

In \cite{myamplitude} we 
have simulated the Blume-Capel model at $D=0.655$  in the high 
temperature phase on lattices of the size $L^3$ with periodic boundary
conditions in all directions and $L \gtrapprox 10 \xi$ for 201 values
of $\beta$.  We have measured the second moment correlation length $\xi_{2nd}$
that we shall define below. The simulation at $\beta=0.3872$, which was our
closest to $\beta_c$,  yielded $\xi_{2nd} = 26.698(7)$.
Fitting these data for $\xi_{2nd}$ with ans\"atze obtained 
by truncating the sequence of correction terms at various order we arrive at
\begin{eqnarray}
\label{xi0}
\xi_{2nd,0,+} &=&  0.2282(2) - 1.8 \times (\nu-0.63002)
                        + 250 \times (\beta_c - 0.387721735) \;\; \nonumber \\
&&  \mbox{using} \;\; t = \beta_c - \beta \;\;
 \mbox{as definition of the reduced temperature} .
\end{eqnarray}
In these fits we have fixed $\nu=0.63002$ and $\beta_c = 0.387721735$ 
(Ref. \cite{mycritical}). 
We have redone the fits with slightly shifted values of 
$\nu$ and  $\beta_c$ to determine the dependence of $\xi_{2nd,0,+}$ on these
input parameters. For simplicity we shall use $t = \beta_c - \beta$ as
reduced temperature also in the following.

In the high temperature phase there is little difference between
$\xi_{2nd}$ and the exponential correlation length $\xi_{exp}$ which
is defined by the asymptotic decay of the two-point correlation function.
Following  \cite{pisaseries}:
\begin{equation}
\lim_{t\searrow 0} \frac{\xi_{exp}}{\xi_{2nd}} = 1.000200(3)  
\;\;
\end{equation}
for the thermodynamic limit of the three-dimensional system. This means that
at the level of our accuracy we can ignore this difference. Note that in the 
following $\xi_{0}$ always refers to $\xi_{2nd,0,+}$, eq.~(\ref{xi0}).

\subsection{Film geometry and boundary conditions}
In the present work we study the thermodynamic Casimir effect 
for systems with film 
geometry. In the ideal case this means that the system has a finite 
thickness $L_0$, while in the other two directions the thermodynamic 
limit $L_1, L_2 \rightarrow \infty$ is taken. In our  Monte Carlo 
simulations we shall study lattices with $L_0 \ll L_1, L_2$ and 
periodic boundary conditions in the $1$ and $2$ directions.  
Throughout we shall simulate lattices with $L_1=L_2=L$.

In the 0 direction we take symmetry breaking boundary conditions. 
In the reduced Hamiltonian of the Blume-Capel model  these can
be implemented by
\begin{equation}
\label{Isingaction2}
H = -\beta \sum_{<xy>}  s_x s_y
  + D \sum_x s_x^2 \;-\; h_1 \sum_{x_0=0,x_1,x_2} s_x 
                   \;-\; h_2 \sum_{x_0=L_0+1,x_1,x_2} s_x \;\; ,
\end{equation}
where $h_1, h_2 \ne 0$ break the symmetry at the surfaces that we
have put on $x_0=0$ and $x_0=L_0+1$. Hence $L_0$ gives the
number of layers in the interior of the film.

In our Monte Carlo simulations we consider the limit of infinitely 
strong surface
fields  $h_1$ and $h_2$, which means that the spins at the surface
are fixed to either $-1$ or $1$, depending on the signs of $h_1$
and $h_2$. Therefore we have implemented in our simulation code
$++$ boundary conditions by setting $s_x=1$ for all $x$ with  $x_0=0$ or
$x_0=L_0+1$ and  $+-$ boundary conditions by setting
$s_x=1$ for all $x$ with  $x_0=0$ and $s_x=-1$ for all $x$ with
$x_0=L_0+1$.
Alternatively, 
these fixed spins could be interpreted as finite surface fields with 
$|h_1|=|h_2|=\beta$ acting on the spins at $x_0=1$ and $x_0=L_0$, 
respectively. 

\section{Observables}

\subsection{Internal energy and free energy}
\label{defineE}
The reduced free energy per area  is defined by
\begin{equation}
\label{fdef1}
f = - \frac{1}{L_1 L_2} \ln Z \;.
\end{equation}
This means that compared with the free energy per area $\tilde f$, 
a factor $k_B T$ is skipped.

Correspondingly we define the energy per area as the derivative of
minus the reduced free energy per area with respect to $\beta$:
\begin{equation}
\label{Edef1}
E = \frac{1}{L_1 L_2} \frac{\partial \ln Z}{\partial \beta} 
  = \frac{1}{L_1 L_2}  \left \langle \sum_{<x,y>} s_x  s_y \right \rangle \; .
\end{equation}
It is straight forward to determine $E$ 
in Monte Carlo simulations.  From the definition of  $E$ follows
\begin{equation}
\label{integrateF}
 f(\beta) = f(\beta_0) - \int_{\beta_0}^{\beta}
                       \mbox{d} \tilde \beta   E(\tilde \beta)   \;\;.
\end{equation}

\subsection{The magnetization profile of films}
\label{mag}
The film is 
invariant under translations in the 1 and 2 direction of the lattice. 
Therefore the magnetization only depends on $x_0$ and we can average 
over $x_1$ and $x_2$:
\begin{equation}
 m(x_0) \;=\; \frac{1}{L^2}  \sum_{x_1,x_2}  \langle s_x  \rangle \;\;.
\end{equation}
Since the film is symmetric for $++$ boundary conditions and 
anti-symmetric for $+-$ boundary conditions under reflections at the middle 
of the film, $m(x_0) = m(L_0-x_0+1)$ for $++$ boundary conditions and 
$m(x_0) = - m(L_0-x_0+1)$ for $+-$ boundary conditions.

\subsection{Second moment correlation length of the films}
\label{xisection}
We have measured the second moment correlation length of the films 
in the 1 and 2 direction of the lattice. To this end we have computed
the connected correlation function of the Fourier transformed field 
\begin{equation}
\label{subtract}
\tilde G(k_1,k_2) \;= \; \langle |\psi(k_1,k_2)|^2   \rangle - 
                   \delta_{(k_1,k_2),(0,0)} L_0 L^2 m^2 
\end{equation}
where $m$ is the  magnetization and the Fourier transformed field
\begin{equation}
 \psi(k_1,k_2) = \frac{1}{\sqrt{L_0 L^2} } 
\sum_x \exp\left(i \frac{2 \pi [k_1 x_1+k_2 x_2]}{L} \right) s_x  \;\;.
\end{equation}
For large $L$ and small $k_1$, $k_2$, the correlation function behaves as
\begin{equation}
\label{behavior}
\tilde G(k_1,k_2) \;= \; 
\frac{C}{4 \sin^2(\pi k_1/L) +4 \sin^2( \pi k_2/L) + \xi_{2nd}^{-2}}  \;\;.
\end{equation}
The second moment correlation length $\xi_{2nd}$ can now be evaluated by 
computing $\tilde G(k_1,k_2)$  for two values of $(k_1,k_2)$ and solving
eq.~(\ref{behavior})  with respect to $\xi_{2nd}^{2}$. In the limit 
$L \rightarrow \infty$ all choices of $(k_1,k_2)$ lead to the same result for
$\xi_{2nd}^{2}$. However, for finite $L$ the deviations from this limit
increase with increasing values of $k_1$ and $k_2$. Therefore, for
$+-$ boundary conditions, we have computed the correlation function at
$(k_1,k_2)=(0,0)$ and $(1,0)$. One gets
\begin{equation}
 \xi_{2nd}^{2} = \frac{\tilde G(0,0)/\tilde G(1,0) -1 }
                      {4 \sin^2(\pi/L) } \;\;.
\end{equation}
In the simulation we have also measured $\tilde G(0,1)$ and have averaged
$\tilde G(1,0)$ and $\tilde G(0,1)$ to reduce the statistical error.

In contrast to $+-$ boundary conditions, for $++$ boundary conditions
there is a finite magnetization at any finite temperature. In order to 
avoid the technical complication of subtracting the magnetization squared
required for $(k_1,k_2)=(0,0)$, eq.~(\ref{subtract}), we have used 
$(k_1,k_2)=(1,0)$ and $(1,1)$ to determine the second moment correlation length
\begin{equation}
 \xi_{2nd}^{2} = \frac{\tilde G(1,0) - \tilde G(1,1) }
                      {[2 \tilde G(1,1)- \tilde G(1,0)] \; 4 \sin^2(\pi/L) }
\;\;.
\end{equation}
In the simulations below we have chosen the lattice size $L$ such that 
the limit $L \rightarrow \infty$ is well approximated.
Hence  $\xi_{2nd}$ is a function of the parameters $\beta$ and $D$
of the model and the thickness $L_0$ of the film.

\section{Finite size scaling}
\label{FFSsection}
The reduced excess free energy of the film behaves as
\begin{equation}
\label{defineh}
f_{ex}(L_0,t)=f_{film}(L_0,t)-L_0 f_{bulk}(t)
\simeq L_0^{-d+1}  h(t [L_0/\xi_0]^{1/\nu}) \;\;,
\end{equation}
where $f_{film}(L_0,t)$ is the reduced free energy per area of the 
film, $f_{bulk}(t)$ the reduced free energy density of the bulk 
system, $h(t [L_0/\xi_0]^{1/\nu})$ is the universal finite size scaling 
function of the excess free energy and $d=3$ is the dimension of the bulk 
system. Here and in the following $\xi_0$ is the amplitude of the 
second moment correlation length of the bulk system in the 
high temperature phase.

Inserting the finite size scaling ansatz~(\ref{defineh}) for the excess
free energy into (\ref{defineFF}) one gets
\begin{eqnarray}
\label{mastermind}
F_{Casimir} &\simeq& - k_B T
\frac{\partial \left[L_0^{-d+1}  h(t [L_0/\xi_0]^{1/\nu})\right]}{ \partial L_0}
\nonumber \\
 &=& - k_B T L_0^{-d} \left[-(d-1) h(t [L_0/\xi_0]^{1/\nu}) +
 \frac{1}{\nu}  t [L_0/\xi_0]^{1/\nu} h'(t [L_0/\xi_0]^{1/\nu}) \right]
 \nonumber \\
           &=& k_B T L_0^{-d} \theta(t [L_0/\xi_0]^{1/\nu})
\end{eqnarray}
where
\begin{equation}
\label{important}
\theta(x) = (d-1) h(x) - \frac{x}{\nu} h'(x) \;
\end{equation}
is the finite size scaling function of the thermodynamic Casimir force and 
$x=t [L_0/\xi_0]^{1/\nu}$. This relation is well known and can be found e.g.
in \cite{Krech}. 

Following the discussion in section III B of ref. \cite{Barber}, taking 
into account leading corrections to scaling one gets
\begin{equation}
\label{hcorrection1}
f_{ex}(L_0,t)=
 L_{0}^{-d+1} \;
\bar{h}(x ,a(D) L_0^{-\omega}) =
L_{0}^{-d+1} h(x) \times (1 + a(D) c(x)
L_0^{-\omega} + ...) 
\end{equation}
and correspondingly for the thermodynamic Casimir force per area
\begin{equation}
\label{hcorrection2}
F_{Casimir} =
  k_B T  L_{0}^{-d} \;
\bar{\theta}(x ,a(D) L_0^{-\omega}) =
k_B T L_{0}^{-d} \; \theta(x) \times (1 + a(D) d(x)
L_0^{-\omega} + ...) \;\;,
\end{equation}
where we have performed the Taylor expansion of $\bar{h}$ 
and $\bar{\theta}$ in their second argument to leading order.
The authors of \cite{VaGaMaDi07,VaGaMaDi08} arrive at a similar expression
as eq.~(\ref{hcorrection2}). Fitting their data, obtained for the Ising model,
they have approximated the function $d(x)$ by a constant. For the improved 
model that we study here $a(D) \approx 0$ holds, which simplifies the 
analysis of our data.

The exponent of the leading correction to scaling takes the value 
$\omega = 0.832(6)$ (Ref. \cite{mycritical}). Furthermore there are subleading 
corrections. Among these, the leading ones come with the exponents 
$\omega' = 1.67(11)$ (Ref. \cite{NewmanRiedel})
and due to the breaking of rotational symmetry 
by the lattice $\omega'' \approx 2$ (Ref. \cite{pisa97}).  
At the level of accuracy
of our data, we can not resolve the individual subleading corrections.
In order to get some estimate of the effect of these corrections on our final
results, we have included a term $c L_0^{-2}$ into the  
ans\"atze~(\ref{xiatbco},\ref{enefit2},\ref{pro2},\ref{DfbcC}) below.

A discussion of corrections caused by the boundaries is given in
section V A of ref. \cite{Barber}. Corrections might arise from irrelevant
surface scaling fields. Furthermore Capehart and Fisher \cite{CaFi76}
have argued that there is an arbitrariness in the definition of the thickness
of the film leading to corrections $\propto L_0^{-1}$.  These two arguments
might be actually unified:
In a real-space Renormalization Group treatment of surface critical phenomena 
one splits the reduced Hamiltonian into a bulk and a surface part. In the
neighborhood of the critical point, one might expand the bulk and 
the surface part of the reduced Hamiltonian into so called scaling fields.
The basic idea is that splitting the reduced Hamiltonian into a 
bulk and a surface part is a priori quite ad hoc. Roughly speaking, 
one might put the contribution for  $ (1-l_s)/2 < x_0 < L_0 + (1 +l_s)/2$ 
of eq.~(\ref{Isingaction}) into the bulk part and the remainder into the 
surface part. This way, the amplitudes of the surface scaling fields 
become functions of $l_s$. Here we do not elaborate what sense can be given to 
non-integer values of $l_s$. The amplitude of the leading irrelevant surface 
scaling field, viewed as a function of $l_s$, might have a zero that we shall 
call $L_s$ in the following. Then this surface scaling field has the RG 
exponent $y_s = -\omega_s = -1$. If there is only one surface scaling field
with the RG exponent $y_s = -1$, corrections  $\propto L_0^{-1}$ can hence
be eliminated by replacing $L_0$ by $L_{0,eff}=L_0+L_s$ in finite size scaling 
laws. 

For the ordinary surface universality class,  
the problem of corrections has been worked out in some detail.
A field theoretical calculation \cite{DiDiEi83} predicts a single
irrelevant scaling field with the RG exponent $y_s = -1$.
These corrections to scaling are related with the extrapolation 
length, which was introduced in the context of mean-field 
theory; See the review \cite{BinderS}.
It is given by the zero of the extrapolated magnetization profile.
The authors of \cite{KiOk85} have employed the concept of the 
extrapolation length in their Monte Carlo study
of the magnetization profile of the three-dimensional 
Ising model on the simple cubic lattice with free boundary conditions,
which belong to the ordinary surface universality class. They have 
simulated various values of the ratio $w$ of the surface
and the bulk coupling. They find that the data for different values of 
$w$ only 
fall nicely on a single scaling curve, when the extrapolation length
that depends on $w$ is properly taken into account. Finally we like 
to mention that there had been attempts to eliminate corrections 
due to the surface by a proper choice of $w$ \cite{krech}.

It is beyond the scope of the present manuscript to check whether 
the result of the field theoretical calculation  \cite{DiDiEi83}  
carries over to the extraordinary surface universality class,
which is relevant for the present study. Our working hypothesis
is that there is only a single irrelevant surface
scaling field with the RG exponent $y_s = -1$ which can 
be accounted for by an effective thickness  $L_{0,eff}$ of the 
film. Furthermore we assume that there are no other 
irrelevant surface scaling fields  with $y_s \gtrapprox - 2$.
The analysis of our precise numerical data for various quantities 
provides a quite non-trivial challenge of this hypothesis.

Finally let us spell out how the effective thickness $L_{0,eff}$
enters into finite size scaling laws.
For the thermodynamic Casimir force one gets
\begin{equation}
\label{mastermindeff}
F_{Casimir} = k_B T L_{0,eff}^{-d} \theta(t [L_{0,eff}/\xi_0]^{1/\nu})
\end{equation}
where both the prefactor $L_{0}^{-d}$ as well as the scaling variable
$x= t [L_{0}/\xi_0]^{1/\nu}$ are replaced by $L_{0,eff}^{-d}$ and 
$x= t [L_{0,eff}/\xi_0]^{1/\nu}$, respectively. 

We also study the finite size scaling behavior of the second moment correlation length
of the film. Taking into account boundary corrections we get
\begin{equation}
 \xi_{2nd,film} = L_{0,eff} X(t [L_{0,eff}/\xi_0]^{1/\nu}) \;\;.
\end{equation}
The magnetization profile at the bulk critical point behaves as
\begin{equation}
\label{magfinite}
 m(x_0) = c \; L_{0,eff}^{-\beta/\nu} \; \psi(z/L_{0,eff})\;\;,
\end{equation}
where  $z=x_0 -L_0/2-1/2$ gives the distance from the middle of the film and
$c$ is a model specific constant that could be fixed by the behavior of the
magnetization or the magnetic susceptibility in the thermodynamic limit.
From scaling relations it follows that $\beta/\nu=(1+\eta)/2$, where 
$\eta=0.03627(10)$ for the three-dimensional Ising universality class 
\cite{mycritical}. Note that the scaling function $\psi(z/L_{0,eff})$ diverges as
$z/L_{0,eff} \rightarrow \pm 1/2$, since the magnetization in the neighborhood of 
the boundary stays finite as $L_0 \rightarrow \infty$ for the boundary conditions
studied here. 

\subsection{Thermodynamic Casimir force and the transfermatrix}
The partition function of the system with fixed boundary conditions 
can be expressed in terms of the eigenvalues of the transfermatrix 
and the overlap of the eigenvectors with the boundary states.
Let us consider a lattice of the size $L_0 \times L^2$, where 
$L$ is large compared with the bulk correlation length but still 
finite. 
We consider the transfermatrix $T$ that acts on vectors that are 
build on the configurations living on $L^2$ slices. We denote the 
eigenvalues of $T$ by $\lambda_{\alpha}$ and the corresponding 
eigenvector by $|\alpha \rangle$, where $\alpha=0,1,2,...,\alpha_{max}$.
The eigenvalues are ordered such that 
$\lambda_{\alpha} \ge \lambda_{\beta}$ for $\alpha < \beta$. 
Note that $T$ commutes with translations, rotations, reflections and with 
the change of the sign of all spins in a slice. Therefore the states 
$|\alpha \rangle$ can be classified according to their momentum, 
the angular momentum, their parity and their behavior under sign-change
of the spins. Note that on the lattice, only a sub-group of the symmetries 
of the continuum is realized. For a detailed discussion of the implications 
of this fact see for example section 3.2 of \cite{ACCH97}, where the 
spectrum of the Ising gauge model in $2+1$ dimensions had been studied. 

Now we can write the partition function of the system with fixed 
boundaries as 
\begin{equation}
\label{transfer}
Z_{b_1,b_2} = \sum_{\alpha} \lambda_{\alpha}^l \; \langle b_1 | \alpha \rangle
                                                \langle b_2 | \alpha \rangle
\;\;,
\end{equation}
where $l=L_0+1$ for our definition of the thickness $L_0$. 
The boundary states $b_{1,2}$ can  be either $+$  or $-$ here.
Note that 
these boundary states are invariant under all symmetries
discussed above except for the sign-change of the spins.  Therefore only
states $|\alpha \rangle$ with zero momentum, zero angular momentum and 
even parity have a non-vanishing overlap  $\langle b | \alpha \rangle$.
Now we can compute the thermodynamic Casimir force per area starting from 
eq.~(\ref{transfer})
\begin{eqnarray}
\frac{1}{k_B T} F_{Casimir} &=&\frac{1}{L^2} \; \frac{\partial}{\partial l} 
 \left[ \ln  Z_{b_1,b_2} - l \ln \lambda_0 \right] \nonumber \\
   &=&
\frac{1}{L^2} \; \frac{\sum_{\alpha} \ln(\lambda_{\alpha}/\lambda_0)\;
(\lambda_{\alpha}/\lambda_0)^l \;
 \langle b_1 | \alpha \rangle \langle b_2 | \alpha \rangle}
{\sum_{\alpha} (\lambda_{\alpha}/\lambda_0)^l \;
 \langle b_1 | \alpha \rangle \langle b_2 | \alpha \rangle}  \;\;,
\end{eqnarray}
where $\lambda_0$ is the largest eigenvalue. Introducing the 
inverse correlation lengths
$1/\xi_{\alpha} = m_{\alpha} = - \ln(\lambda_{\alpha}/\lambda_0)$
we get
\begin{equation}
\label{TCasimir}
\frac{1}{k_B T} F_{Casimir} = 
- \frac{1}{L^2} \; \frac{\sum_{\alpha} m_{\alpha}  \exp(-m_{\alpha} l) \;
  \langle b_1 | \alpha \rangle \langle b_2 | \alpha \rangle}  
  {\sum_{\alpha} \exp(-m_{\alpha} l)  \;
 \langle b_1 | \alpha \rangle \langle b_2 | \alpha \rangle} \;\;.
\end{equation}
This equation proves that for $b_1=b_2$ the thermodynamic Casimir 
force takes negative values. In the high temperature phase, in the 
zero momentum sector, the second largest eigenvalue $\lambda_1$ is well 
separated from larger eigenvalues. Therefore the behavior of the thermodynamic
Casimir force for $l \gg \xi_1=\xi=1/m$, which corresponds to large values of
the scaling variable $x$, is given by
\begin{eqnarray}
\tilde \theta(m l) &\approx& \frac{l^3}{k_B T} F_{Casimir} \approx
- \frac{m l^3 \exp(-m l) \;
 \frac{1}{L^2}  \langle b_1 | 1 \rangle \langle b_2 | 1 \rangle}
  {
 \langle b_1 | 0 \rangle \langle b_2 | 0 \rangle +
 \exp(-m l) \; \langle b_1 | 1 \rangle \langle b_2 | 1 
\rangle}  \nonumber \\
&\approx &
- m^3 l^3 \exp(-m l) \; \frac{1}{m^2 L^2} 
 \frac{\langle b_1 | 1 \rangle \langle b_2 | 1 \rangle}
      {\langle b_1 | 0 \rangle \langle b_2 | 0 \rangle} \;\;.
\end{eqnarray}
The finite size scaling behavior~(\ref{mastermind}) of the thermodynamic  
Casimir force implies that
\begin{equation}
C(b) = \frac{1}{m L} \frac{\langle b | 1 \rangle}
                          {\langle b | 0 \rangle}
\end{equation}
has a finite scaling limit.  The state $|0 \rangle$ is symmetric 
under the global transformation $s_x \rightarrow -s_x$ for all $x$ in 
a slice. Instead,  $|1 \rangle$ is 
anti-symmetric and therefore $C=C(+) = - C(-)$.  It follows
\begin{equation} 
\label{XZX}
\tilde \theta_{++}(m l) = - \tilde \theta_{+-}(m l) = 
- C^2 \; m^3 l^3 \exp(-m l)
\end{equation}
for sufficiently large values of $m l$. Since 
$x = t [l/\xi_0]^{1/\nu} \simeq (m l)^{1/\nu}$ it follows
\begin{equation}
\label{scalinghigh}
\theta_{++}(x) = - \theta_{+-}(x) = - C^2  x^{3 \nu} \exp(-x^{\nu})
\end{equation} 
for sufficiently large values of $x$.  In the low temperature phase, 
the situation is more complicated. Also here, for finite $L$ the 
state $|0 \rangle $ is symmetric under $s_x \rightarrow -s_x$, while 
$|1 \rangle$ is anti-symmetric. The corresponding correlation length 
$\xi_t = -1/\ln(\lambda_1/\lambda_0)$ is the so called tunneling 
correlation length. It diverges as $\xi_t \propto \exp(\sigma L^2)$  
in the limit $L \rightarrow \infty$, where $\sigma$ is the interface 
tension. It is characteristic for the low temperature phase, and
a consequence of spontaneous symmetry breaking that pairs of eigenvalues,
where one is symmetric and the other anti-symmetric under 
$s_x \rightarrow -s_x$, become degenerate in the limit $L \rightarrow \infty$.
The bulk correlation length in the low temperature phase is given by
$\xi=- \lim_{L\rightarrow \infty} 1/\ln(\lambda_{2}/\lambda_{0})=
- \lim_{L\rightarrow \infty} 1/\ln(\lambda_{3}/\lambda_{0})$. Taking into
account the states $\alpha=0, 1, 2$ and $3$ we get 
\begin{equation}
\label{lowtrans}
\frac{1}{k_B T} F_{Casimir} \approx - \frac{1}{L^2} 
\frac{
      m_2 \exp(-m_2 l) \langle b_1| 2 \rangle \langle b_2| 2 \rangle
     +m_3 \exp(-m_3 l) \langle b_1| 3 \rangle \langle b_2| 3 \rangle
      }{\langle b_1| 0 \rangle \langle b_2| 0 \rangle +
        \exp(-m_t l) \langle b_1| 1 \rangle \langle b_2| 1 \rangle } \;\;,
\end{equation}
where we have skipped the contribution of $\alpha=1$ in the numerator, since 
$m_t$ vanishes in the limit $L \rightarrow \infty$. Furthermore, we have 
skipped the contributions of $\alpha=2$ and $3$ in the denominator, 
since for $m_2 l, m_3 l \gg 1$ they are small compared with those of 
$\alpha=0$ and $1$. For $+-$ boundary conditions 
$\langle +| \alpha \rangle \langle -| \alpha \rangle$ is positive for 
states that are symmetric and negative for states that are anti-symmetric
under the spin-flip. Therefore both in the numerator and the denominator
there is a cancellation  between the two terms. Extracting useful information 
from eq.~(\ref{lowtrans}) would require detail knowledge of the approach 
of $m_t$, $m_2$, $m_3$ and the overlap amplitudes to the limit 
$L\rightarrow \infty$.

On the other hand for $++$ boundary conditions 
$\langle +| \alpha \rangle \langle +| \alpha \rangle$ is positive for any 
$\alpha$. Therefore in eq.~(\ref{lowtrans}) the two terms in the numerator
and the denominator add up. In the limit $L \rightarrow \infty$, where
$m_t = 0$ and $m=m_2=m_3$ we get a result analogous to eq.~(\ref{XZX}). 
We only have to 
notice that in the definition of the scaling variable $x$ the amplitude 
$\xi_{0,+}$ of the correlation length in the high temperature phase enters.
Therefore taking into account the universal amplitude ratio 
$\xi_{0,+}/\xi_{0,-}=1.901(14)$  (Ref. \cite{myamplitude}) for the exponential 
correlation length we get
\begin{equation}
\label{scalinglow}
\theta_{++}(x) = - \bar{C}^2  [-1.901(14) x]^{3 \nu} \exp(-[-1.901(14) x]^{\nu})
\end{equation}
for sufficiently small values of $x$ in the low temperature phase.
For a discussion of the spectrum and the symmetry properties
of the eigenvectors of the transfermatrix see e.g. \cite{Klessinger}.
Eqs.~(\ref{scalinghigh}, \ref{scalinglow}) had been derived before 
by using the  de Gennes-Fisher local-functional method, see eq.~(6) of ref.
\cite{BoUp08}. Exact results for the Ising strip \cite{EvSt94} and mean-field 
theory \cite{Krech97} confirm the exponential decay of $\theta_{++}(x)$ 
for large $|x|$.

\section{Monte Carlo algorithms} 
\subsection{$++$ boundary conditions}
In the case of $++$ boundary conditions we have used a
hybrid of a cluster update and a local heat bath algorithm \cite{BrTa}. 
The cluster algorithm can only change the sign of the spins. Therefore
local heat bath updates are needed to get an ergodic algorithm.
For the cluster algorithm, we have used the same probability to 
freeze or delete a link $<xy>$ as it is used in the original Swendsen-Wang
\cite{SW} algorithm:
\begin{equation}
\label{probdel}
p_d(s_x s_y) = \mbox{min}[1,\exp(-2 \beta s_x s_y)] \;\;.
\end{equation}
Links are deleted with the probability $p_d(s_x s_y)$, otherwise they 
are frozen. A cluster is a set of sites that is connected by frozen links.
In the following we mean by ``flipping a cluster'' that the sign of all 
spins $s_x$, where the site $x$ belongs to the cluster,  
is changed (``flipped'').
In one step of the Swendsen-Wang cluster algorithm, the lattice is completely
decomposed into clusters. A cluster is then flipped with the 
probability $1/2$. In contrast, in the case of the Wolff single cluster 
algorithm \cite{Wolff}, one site of the lattice is chosen randomly. 
Then only the cluster that contains this site is constructed. This 
cluster is flipped with probability 1.  Here we have to deal with the 
boundaries. For links $<xy>$, where either $x$ or $y$ belongs to the boundary 
we shall apply the same freeze or delete probability~(\ref{probdel}) as for 
links $<xy>$, where none of the two sites belongs to the boundary.  Since
spins on the boundary are fixed to one, clusters that contain 
sites on the boundary can not be flipped.  Motivated by this fact, 
we have flipped all clusters with probability one that do not include
sites on the boundary. In practice this is
done in the following way: First we compute all clusters that include sites
on the boundary. Then all  spins on sites that do not belong to these
clusters are flipped.

With the local heat bath algorithm we run through the lattice in typewriter
fashion. Running through the lattice once is called one ``sweep'' in the 
following.
One cycle of the hybrid algorithm is composed of two sweeps of the local
heat bath algorithm followed by one cluster update as discussed above.
At the bulk critical point the integrated autocorrelation time of the 
energy is $\tau_{int,E} \approx 3$ in units of update cycles 
for a lattice of the size $L_0=32$, $L_1=L_2=128$.
The integrated autocorrelation times for $\tilde G(1,0)$ and $\tilde G(1,1)$ 
are smaller.

\subsection{$+-$ boundary conditions}
We could not use the program written for the 
$++$ boundary conditions for the $+-$ boundary conditions,
since it relies on the 
fact that all spins that belong to clusters that  include sites on the 
boundary are equal to $+1$. For simplicity we therefore have 
used a local Metropolis algorithm that was implemented by using the 
multispin coding technique \cite{multispin}.
Details of our implementation can be found in \cite{mycritical}. 
In \cite{mycritical} we have found a performance gain of our Metropolis
update using the multispin coding technique of about a factor of ten 
compared with the heat bath algorithm, implemented in a standard way.

Likely, for small values of $L_0$ the local Metropolis algorithm
implemented by using the multispin coding technique outperforms the 
hybrid of local heat bath and cluster algorithm in the case of $++$ 
boundary conditions.  For lack of time we did not check this.

In the low temperature phase, for $+-$ boundary conditions rather large 
autocorrelations arise. These are due to fluctuations of the interface 
between the $+$ and the $-$ phase. As discussed in \cite{myPRL} standard
cluster algorithms are not suitable to overcome this problem. 
Unfortunately, the algorithm discussed in \cite{myPRL} only works well 
in the Ising limit.  

In all our simulations we have used the  SIMD-oriented Fast
Mersenne Twister algorithm \cite{twister} as random number generator.

\section{Simulations at the bulk critical point}
Here we focus on the finite size scaling behavior of various quantities 
at the bulk critical point. This way we accurately compute $L_s$, which 
characterizes the corrections caused by the boundary conditions. To this 
end we have performed two sets  of simulations.
First we have simulated films of the size $L_0 \times L^2$ to determine 
the second moment correlation length in 1 and 2 directions, the energy 
per area of the films and the magnetization profile.
Then we computed the differences  
\begin{equation}
 \Delta f(L_0,\beta_c) = f(L_0+1/2,\beta_c)- f(L_0-1/2,\beta_c)
\end{equation}
of free energies per area, where  $L_0+1/2$ and $L_0-1/2$ assume integer 
values. To this end, we have simulated a lattice with $L_0-1/2$ complete 
layers and one incomplete layer. $\Delta f(L_0,\beta_c)$ is then given 
by the free energy required to add a single site to this incomplete layer.
For details of the method see \cite{yet}.  

\subsection{Correlation length and  energy per area at the bulk 
            critical point}
\label{eneatbc}
For both $+-$ and $++$ boundary conditions we have simulated lattices of the
thicknesses $L_0=6, 7, 8,...,26, 28, 30, 32$. Throughout we have used 
$L = 4 L_0$. At the bulk critical point,
the correlation length of films with $++$ boundary conditions
is $\xi_{2nd} \approx 0.13 L_0$ and for $+-$ boundary conditions
$\xi_{2nd} \approx 0.21 L_0$, as we shall see below. Therefore this choice 
of $L$ is sufficient to get a good approximation of the limit 
$L \rightarrow \infty$. Throughout we have performed 
$100 \; 000 \; 000$ update cycles for $++$ boundary conditions and 
$64 \times 5 \; 000 \; 000$ measurements for $+-$ boundary conditions.
In the case of $+-$ boundary conditions up to 18 Metropolis sweeps were 
performed for each measurement.  In total the simulations took one year and
1.5 years on one core of a Quad-Core AMD Opteron(tm) Processor 2378 running
at 2.4 GHz for $++$ and $+-$ boundary conditions, respectively.

We have fitted the second moment correlation length at the critical 
point of the bulk system with the ansatz
\begin{equation}
\label{xiatbc}
 \xi_{2nd} = c \; (L_0 + L_s)
\end{equation}
and to check for the possible effect of subleading corrections
\begin{equation}
\label{xiatbco}
\xi_{2nd} = c \; (L_0 + L_s) \times (1 + b \; (L_0+L_s)^{-2})  \;\;.
\end{equation}
For $++$ boundary conditions, fitting with ansatz~(\ref{xiatbc}) we get
for $L_{0,min}=12$ the results $c= 0.1303(2)$, $L_s=1.89(3)$
and $\chi^2/$d.o.f.$=0.83$. 
In this fit we have taken all data with $L_0 \ge L_{0,min}$ into account.
Using instead the ansatz~(\ref{xiatbco}) we get for $L_{0,min} =6$ the results
$c=0.1303(2)$,  $L_s=1.89(4)$ and $\chi^2/$d.o.f.$=0.94$. 

For $+-$ boundary conditions, fitting with ansatz~(\ref{xiatbc}) we get  
for $L_{0,min}=14$ the results $c= 0.2111(3)$, $L_s=2.01(3)$ 
and $\chi^2/$d.o.f.$=1.97$. 
Using instead the ansatz~(\ref{xiatbco}) we get for $L_{0,min} =8$ the results
$c= 0.2119(4)$,  $L_s=1.81(6)$ and $\chi^2/$d.o.f.$=2.08$. In both cases, 
the $\chi^2/$d.o.f. does not further decrease with increasing $L_{0,min}$.

We conclude that the results obtained for $L_s$ for the $++$ and the $+-$
boundary conditions are both consistent with $L_s \approx 1.9$. We have 
checked that the error of $\beta_c$ can be safely ignored. 

Next we have fitted the excess energy per area at the bulk critical point 
with the ansatz 
\begin{equation}
\label{enefit}
 E_{ex}(L_0,\beta_c) = B +  a \; (L_0 + L_s)^{-2 + 1/\nu}  \;\;,
\end{equation}
where we have used $E_{bulk}(\beta_c) = 0.602111(1)$ (Ref. \cite{myamplitude}) 
to compute
$E_{ex}(L_0,\beta_c)$ and we  have fixed $\nu=0.63002$ (Ref. \cite{mycritical}). 
The parameters of the fit are $B$, $a$ and $L_s$. Note that $B$ corresponds
to a correction of the analytic background caused by the boundaries 
that only depends on the local properties of the system at the boundaries
and therefore takes the same value for $++$ and $+-$ boundary conditions.
In order to estimate errors due to subleading corrections we have also fitted
with
\begin{equation}
\label{enefit2}
 E_{ex}(L_0,\beta_c) = B +  a \; (L_0 + L_s)^{-2 + 1/\nu}  
                   \times (1 + c \; (L_0 + L_s)^{-2}) \;\;,
\end{equation}
where we have included quadratic corrections.

For $++$ boundary conditions we get with the ansatz~(\ref{enefit}) 
for $L_{0,min}=8$ the results
$B=7.1893(3)$, $a=-8.045(1)$, $L_s=1.915(2)$ and $\chi^2/$d.o.f. $=0.79$.
Using the ansatz~(\ref{enefit2}) and $L_{0,min}=6$ we get $B=7.1888(2)$, 
$a=-8.042(1)$, $L_s=1.905(1)$ and $\chi^2/$d.o.f. $=0.96$.

Instead, for $+-$ boundary conditions we get using the ansatz~(\ref{enefit}) 
for $L_{0,min}=13$ the results $B=7.1947(4)$, $a=-12.207(2)$, $L_s=1.966(3)$
and  $\chi^2/$d.o.f. $=0.60$.   
Using ansatz~(\ref{enefit2}) we get for $L_{0,min}=8$ the results 
$B=7.1864(5)$, $a=-12.156(3)$, $L_s=1.830(6)$ and  $\chi^2/$d.o.f. $=0.53$.
The results of the two ans\"atze~(\ref{enefit},\ref{enefit2}) differ by 
several standard deviations, indicating that the systematical error due 
to corrections to scaling is clearly larger than the statistical one.
Here we try to estimate this error from the difference between the 
results of the two  ans\"atze~(\ref{enefit},\ref{enefit2}).  Furthermore
we have redone the fits above using shifted values for the input parameters
$E_{bulk}(\beta_c)$ and $\nu$ to estimate the effect of their uncertainty 
on our results. 
In particular we find that by using $\nu=0.63012$ instead of $\nu=0.63002$ 
the values of our fitparameters shift considerably. E.g. for $++$ boundary 
conditions and $L_{0,min}=8$ using ansatz~(\ref{enefit})  we get 
 $B=7.1912(3)$, $a=-8.040(1)$, $L_s=1.909(2)$ and $\chi^2/$d.o.f. $=0.78$.
Taking into account the results of both $++$ and $+-$ boundary 
conditions we arrive at
\begin{eqnarray}
 B&=&7.189(6)  \\
 L_s &=& 1.9(1)  \\
 a_{++} &=&  -8.04(1) \\
 a_{+-} &=& -12.18(3) \;\;,
\end{eqnarray}
where we have taken the error mainly from the difference between the two 
different ans\"atze for the $+-$ boundary conditions.  We notice that the 
result obtained for $L_s$ is fully consistent with that obtained from the 
analysis of the second moment correlation length above.

\subsection{The magnetization profile at the critical point}
In order to determine the constant $L_s$ we have studied the magnetization 
at $z=x_0-(L_0+1)/2=0$, i.e. in the middle of the film, for $++$ boundary conditions. 
In the case of odd $L_0$ we did
use directly the value of the magnetization at $z=0$. In the case of even $L_0$
we extrapolated the values of $m$ at $z=3/2$ and $z=1/2$ to $z=0$, assuming a
quadratic dependence on $z$. For example for $L_0=24$, $25$, $26$, $28$, $30$, 
and $32$ we get $\left. m\right |_{z=0} =0.248488(6)$, $0.243670(4)$, $0.239111(4)$, 
$0.230695(4)$, $0.223091(4)$, and $0.216181(4)$, respectively.

Following eq.~(\ref{magfinite}), we have fitted our data with the ansatz
\begin{equation}
\label{pro1}
\left . m \right |_{z=0} = C_m \; (L_0 + L_s)^{-\beta/\nu}
\end{equation}
where $C_m$ and $L_s$ are the parameters of the fit. Note that $\beta/\nu=(1+\eta)/2$
follows from scaling relations among the critical exponents. In our fits,
we have fixed $\eta=0.03627$ (Ref. \cite{mycritical}).
In order to check for the effect of possible corrections, we have used in 
addition
\begin{equation}
\label{pro2}
\left . m \right |_{z=0} = C_m \; (L_0 + L_s)^{-\beta/\nu} 
            \times (1 + c \; (L_0 + L_s)^{-2}) \;\;.
\end{equation}
Fitting with the ansatz~(\ref{pro1}) we find that the result for $L_s$ 
is slowly decreasing with an increasing minimal thickness $L_{0,min}$ that
is included into the fit. For $L_{0,min}=20$ we find that $\chi^2/$d.o.f. 
is still larger than two.  For $L_{0,min}=24$ we get $C_m=1.34250(10)$, 
$L_s=1.937(4)$ and $\chi^2/$d.o.f.$=0.34$.  We have redone the fit with
$\eta=0.03637$ instead of the central value $\eta=0.03627$. We find that the 
effect on $C_m$ and $L_s$ is much less than the statistical errors quoted 
above. Fitting with the ansatz~(\ref{pro2}) we find for $L_{0,min}=16$
the results $C_m=1.34171(17)$, $L_s=1.867(12)$ and $\chi^2/$d.o.f.$=0.55$.
Also here we find that the error due to the uncertainty of $\eta$ is 
small compared with the statistical error quoted. Our results for $L_s$ 
are in very good agreement with those obtained above.

\begin{figure}
\begin{center}
\includegraphics[width=13.3cm]{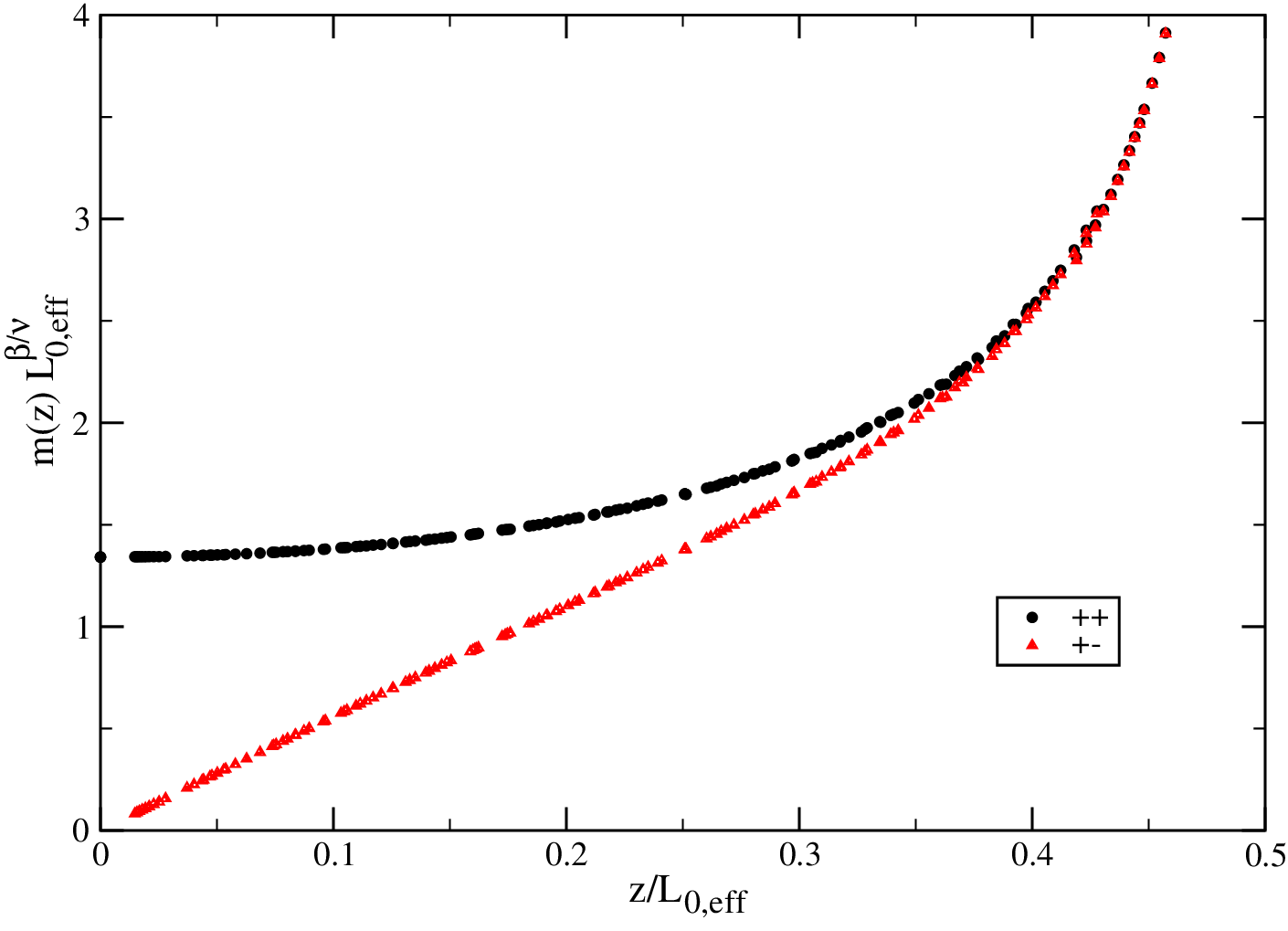}
\caption{\label{profileplot}
We plot $L_{0,eff}^{\beta/\nu} m(z)$
as a function of $z/L_{0,eff}$, where $z=x_0-(L_0+1)/2$ gives the distance 
from the middle of the film. The effective thickness of the 
film is $L_{0,eff}=L_0+L_s$ using $L_s=1.9$.  For $++$ and $+-$ 
boundary conditions, data for films with $L_0 \ge 16$ are used.
}
\end{center}
\end{figure}

Finally in figure \ref{profileplot} we plot 
$L_{0,eff}^{\beta/\nu} m(z)$ as a function of $z/L_{0,eff}$ 
using $L_s=1.9$ and  $\eta=0.03627$ for $++$ and $+-$ boundary conditions.
To this end we have used all thicknesses available with $L_0 \ge 16$. 
The statistical errors are much smaller than the symbols that are used.
For $z/L_{0,eff} \lessapprox 0.4$ the points fall nicely on unique curves for 
$++$ and $+-$ boundary conditions, respectively. For larger values of 
$z$ a small 
scattering of the data can be observed. As the boundary is approached, 
this means $z \rightarrow 1/2$, the curves for $++$ and $+-$ boundary 
conditions fall on top of each other.

\subsection{Casimir force at the critical point}

\label{casimir0}
We have computed
\begin{equation}
 \Delta f(L_0,\beta_c) = f(L_0+1/2,\beta_c)- f(L_0-1/2,\beta_c)
\end{equation}
using the algorithm discussed in ref. \cite{yet}. 
We have simulated $++$ and $+-$ boundary conditions on lattices 
of the thicknesses $L_0=6.5$, $7.5$, $8.5$, $9.5$, $10.5$, $11.5$, 
$12.5$, $13.5$, $15.5$, $19.5$, $23.5$, $27.5$, $31.5$ and $39.5$.
For all these simulations, we have used 
$L \approx 8 L_0$. We have checked that this is sufficient to avoid
finite $L$ corrections.
These simulations took in total about 10 month of CPU-time on
one core of a Quad-Core AMD Opteron(tm) Processor 2378 running at 2.4 GHz.
As update we have used  the local heat bath algorithm.
 For lack of time and the still moderate amount of CPU time that was spent here,
we made no effort to implement cluster updates or to implement the method using 
the multispin coding technique.

We have fitted our data  with the ans\"atze
\begin{equation}
\label{Dfbc}
 \Delta f(L_0,\beta_c) = f_{bulk}(\beta_c) - \theta(0) \; (L_0+L_s)^{-3}
\end{equation}
and in order to check for the effect of subleading corrections to scaling
\begin{equation}
\label{DfbcC}
\Delta f(L_0,\beta_c) = f_{bulk}(\beta_c) -
\theta(0) \; (L_0+L_s)^{-3}  \times (1 + c \; (L_0+L_s)^{-2}) \;\;.
\end{equation}

Fitting with the ansatz~(\ref{Dfbc}) we get for the $++$ boundary conditions 
and $L_{0,min}=11.5$ the results
$f_{bulk}(\beta_c) =- 0.0757368(3)$, $\theta(0) = - 0.815(10)$, $L_s=1.86(5)$
and $\chi^2/$d.o.f.$= 0.36$. 
Using the ansatz~(\ref{DfbcC}) and $L_{0,min}=6.5$ we get the results
$f_{bulk}(\beta_c) =-0.0757370(2)$, $\theta(0) = - 0.824(5)$,  $L_s=1.91(4)$
and $\chi^2/$d.o.f.$= 0.51$.

Fitting with the ansatz~(\ref{Dfbc}) we get for the $+-$ 
boundary conditions and $L_{0,min}=11.5$
the results $f_{bulk}(\beta_c) = -0.0757368(2)$, 
$\theta(0) = 5.617(16)$,  $L_s= 1.930(13)$ and $\chi^2/$d.o.f.$=1.11$. 
Using the ansatz~(\ref{DfbcC}) and $L_{0,min}=6.5$ we get the results
$f_{bulk}(\beta_c) = -0.0757368(2)$, $\theta(0) =5.610(14)$,  $L_s=  1.912(17)$
and $\chi^2/$d.o.f.$= 0.81$. 

We notice that  the results for $f_{bulk}(\beta_c)$ obtained from the 
two different boundary conditions are consistent. We conclude
\begin{equation}
f_{bulk}(\beta_c)  = -0.0757368(4)  \;\;.
\end{equation}
Also the values for $L_s$  obtained here are fully consistent with the
estimate $L_s=1.9(1)$ found above. As our result for the finite size scaling 
functions at the critical point of the bulk system we quote
\begin{eqnarray}
\label{thetapp}
\theta_{++}(0) &=& - 0.820(15) \\
\label{thetapm}
\theta_{+-}(0) &=&\phantom{+} 5.613(20)   \;\; .
\end{eqnarray}
Also here we have checked that the uncertainty of $\beta_c$ can be 
safely ignored. For a comparison of these results with previous ones
given in the literature, see table \ref{Camplitude} below.

\section{Numerical results for the Casimir force in a large range of 
temperatures}
Here we compute the Casimir force using the method discussed by Hucht 
\cite{Hucht}. The details of the implementation are similar to 
\cite{myXY}, where we have studied the thermodynamic Casimir force 
for films with free boundary conditions in the three dimensional 
XY universality class.

We have simulated the model for both types of boundary conditions 
and the thicknesses  $L_0=8, 9, 16, 17, 32$ and $33$ for a large 
number of $\beta$-values in the neighborhood of the critical point. 
In tables \ref{statpm} and \ref{statpp} we give the $\beta$-values at 
which we have simulated and the statistics of our runs for the $+-$ and 
the $++$ boundary conditions, respectively.  In the case of $+-$ 
boundary conditions we also give the lattice size $L$ that was used.
Since for $+-$ boundary conditions
the correlation length is increasing with increasing $\beta$ also 
$L$ has to increase with increasing $\beta$.  
In contrast, for $++$ boundary conditions, the correlation length stays 
rather small for all temperatures. It has a maximum quite close to the 
critical point. Therefore we have used $L=32$  for $L_0=8$, $9$,
$L=64$  for $L_0=16$, $17$  and $L=128$  for $L_0=32$, $33$ at all  values
of $\beta$, where we have simulated at.

We have measured the energy per area. Using these data we have computed
\begin{equation}
\Delta E(L_0,\beta)=E(L_0+1/2,\beta)-E(L_0-1/2,\beta)-E_{bulk}(\beta) \;\;.
\end{equation} 
The value for the energy density of the bulk system $E_{bulk}(\beta)$ 
is taken from simulations of $L^3$ or $2 L \times L^2$ lattices
with periodic boundary conditions in all three directions. The linear lattice 
size $L$ is taken sufficiently large to avoid significant finite size 
effects.  For most values of $\beta$ simulated here we have also a direct 
measurement of $E_{bulk}(\beta)$.  In a small neighborhood of $\beta_c$ 
we have used instead the result of a fit with the ansatz 
\begin{equation}
\label{critical2}
 E_{bulk}(\beta) = E_{ns} + C_{ns} (\beta-\beta_c)
             + a_{\pm} |\beta-\beta_c|^{1-\alpha}
             + d_{ns} (\beta-\beta_c)^2
             + b_{\pm} |\beta-\beta_c|^{2-\alpha} \;\;.
\end{equation}
For a discussion see section IV A of \cite{myamplitude}. Throughout the 
statistical error of $E_{bulk}(\beta)$ is clearly smaller than that of 
$E(L_0+1/2,\beta) - E(L_0-1/2,\beta)$.  Also the systematical error caused
by the interpolation with the ansatz ~(\ref{critical2}) can be safely 
ignored here.

In order to obtain $\Delta f_{ex}$ we have numerically integrated
$\Delta E_{ex}$ using the trapezoidal rule:
\begin{equation}
\label{integration}
-\Delta f_{ex}(\beta_n) \approx \sum_{i=0}^{n-1} \frac{1}{2} (\beta_{i+1}-\beta_i)
   \left(\Delta E_{ex}(\beta_{i+1}) + \Delta E_{ex}(\beta_{i}) \right)
\end{equation}
where $\beta_{i}$ are the values of $\beta$ we have simulated at. They
are ordered such that $\beta_{i+1} > \beta_i$ for all $i$. The starting 
point of the integration $\beta_0$ is chosen such that 
$\Delta E_{ex}(\beta_{0}) =0$  within the statistical error.

The estimate obtained from the integration is affected by statistical
and systematical errors. The statistical one can be easily computed, since
the $\Delta E_{ex}$ are obtained from independent simulations:
\begin{eqnarray}
 \epsilon^2 (-\Delta f_{ex}(\beta_n))
 &=& \frac{(\beta_{1} - \beta_{0})^2}{4} \epsilon^2 [\Delta E_{ex}(\beta_0)]
  + \frac{(\beta_{n} - \beta_{n-1})^2}{4} \epsilon^2 [\Delta E_{ex}(\beta_n)]
  \nonumber \\
 &+& \sum_{i=1}^{n-1} \frac{(\beta_{i+1} - \beta_{i-1})^2}{4}
      \epsilon^2 [\Delta E_{ex}(\beta_i)]
\end{eqnarray}
where $\epsilon^2$ denotes the square of the statistical error.

In order to estimate the error due to the finite step size
$\beta_{i+1}-\beta_{i}$ we have redone the integration, skipping every
second value of $\beta$; i.e. doubling the step size.  We find that
the finite step size errors are at most of the size of the statistical ones.

\begin{table}
\caption{\sl \label{statpm} Statistics of our runs for the
$+-$ boundary conditions. In the first column we give the thickness 
that is considered, where for example $L_0=8.5$ means that we have simulated 
films of the thicknesses $L_0=8$ and $9$. In the second column we give the 
linear extension $L$ of the lattice in 1 and 2 direction. We have simulated 
at $\beta_i=\beta_{min}+i \Delta \beta$ in the interval 
$[\beta_{min},\beta_{max}]$.  In the last column we give the number of 
measurements for each of the simulations.
}
\begin{center}
\begin{tabular}{rrlllr}
\hline
\mc{1}{c}{$L_0$}  & \mc{1}{c}{$L$} &  \mc{1}{c}{$\beta_{min}$}  & 
\mc{1}{c}{$\beta_{max}$}  & \mc{1}{c}{$\Delta \beta$} &\mc{1}{c}{stat} \\
\hline
 8.5   &   32      &  0.25           & 0.325         & 0.005 & 200 000 \\
 8.5   &   32      &  0.33           & 0.348         & 0.002 & 200 000 \\
 8.5   &   32      &  0.35           & 0.38        & 0.001   & 200 000 \\
 8.5   &   32      &  0.381           & 0.385        & 0.001 & 300 000 \\
 8.5   &   64      &  0.385          & 0.43         & 0.001  & 150 000 \\
 8.5   &   96      &  0.43           & 0.46         & 0.002  & 100 000 \\
 8.5   &  128      &  0.46           & 0.5          & 0.002  & 100 000 \\
 8.5   &  256      &  0.505          & 0.56         & 0.005  & 100 000 \\
\hline
 16.5  &  64       & 0.34            & 0.348        & 0.002  &200 000  \\  
 16.5  &  64       & 0.35            & 0.384        & 0.001  &200 000 \\  
 16.5  &  64       & 0.385           & 0.395        & 0.0005 &200 000 \\ 
 16.5  & 128       & 0.395           & 0.41         & 0.001  &100 000 \\
 16.5  & 256       & 0.412           & 0.42         & 0.002  &100 000 \\
 16.5  & 512       & 0.422           & 0.43         & 0.002  &100 000 \\ 
 16.5  & 512       & 0.44            & 0.44         & 0.01   &100 000 \\ 
\hline
 32.5  & 128       & 0.36            & 0.355        & 0.005  &1 000 000 \\
 32.5  & 128       & 0.365           & 0.368        & 0.001  &1 000 000 \\
 32.5  & 128       & 0.369           & 0.3875       & 0.0005 &1 000 000 \\
 32.5  & 128       & 0.3875          & 0.39125      & 0.00025&1 000 000 \\
 32.5  & 256       & 0.3915          & 0.395        & 0.0005 & 250 000 \\
\hline
\end{tabular}
\end{center}
\end{table}

\begin{table}
\caption{\sl \label{statpp} Statistics of our runs for the
$++$ boundary conditions.  The notation is the same as in the 
previous table for $+-$ boundary conditions. Here we have used
$L=4 (L_0-1/2)$ for all values of $\beta$.   
}
\begin{center}
\begin{tabular}{rlllr}
\hline
$L_0$ & $\beta_{min}  $  & $\beta_{max}$ & $\Delta \beta$  &   stat \\
\hline
 8.5 &  0.25           & 0.295         &   0.005         &  5 000 000 \\
 8.5 &  0.3            & 0.348         &   0.002         &  5 000 000 \\
 8.5 &  0.35           & 0.358         &   0.002         & 10 000 000 \\
 8.5 &  0.36           & 0.378         &   0.001         & 10 000 000 \\
 8.5 &  0.379          & 0.395         &   0.0005        & 10 000 000 \\
 8.5 &  0.396          & 0.409         &   0.001         & 10 000 000 \\
 8.5 &  0.41           & 0.43          &   0.002         & 10 000 000 \\
\hline
16.5 &  0.31           & 0.33          & 0.01            & 10 000 000 \\
16.5 &  0.34           & 0.352         & 0.002           & 10 000 000 \\
16.5 &  0.354          & 0.379         & 0.001           & 10 000 000 \\
16.5 &  0.38           & 0.382         & 0.0005          & 10 000 000 \\
16.5 &  0.3825         & 0.39225       & 0.00025         & 10 000 000 \\
16.5 &  0.393          & 0.399         & 0.001           & 10 000 000 \\
16.5 &  0.4            & 0.406         & 0.002           & 10 000 000 \\
\hline
32.5 &  0.37           & 0.375         & 0.001           & 10 000 000 \\
32.5 &  0.376          & 0.3795        & 0.0005          & 10 000 000 \\
32.5 &  0.38           & 0.3856        & 0.0002          & 10 000 000 \\
32.5 &  0.3858         & 0.3889        & 0.0001          & 10 000 000 \\
32.5 &  0.389          & 0.3918        & 0.0002          & 10 000 000 \\
32.5 &  0.392          & 0.3945        & 0.0005          & 10 000 000 \\
32.5 &  0.395          & 0.396         & 0.001           & 10 000 000 \\
\hline
\end{tabular}
\end{center}
\end{table}

\begin{figure}
\begin{center}
\includegraphics[width=13.3cm]{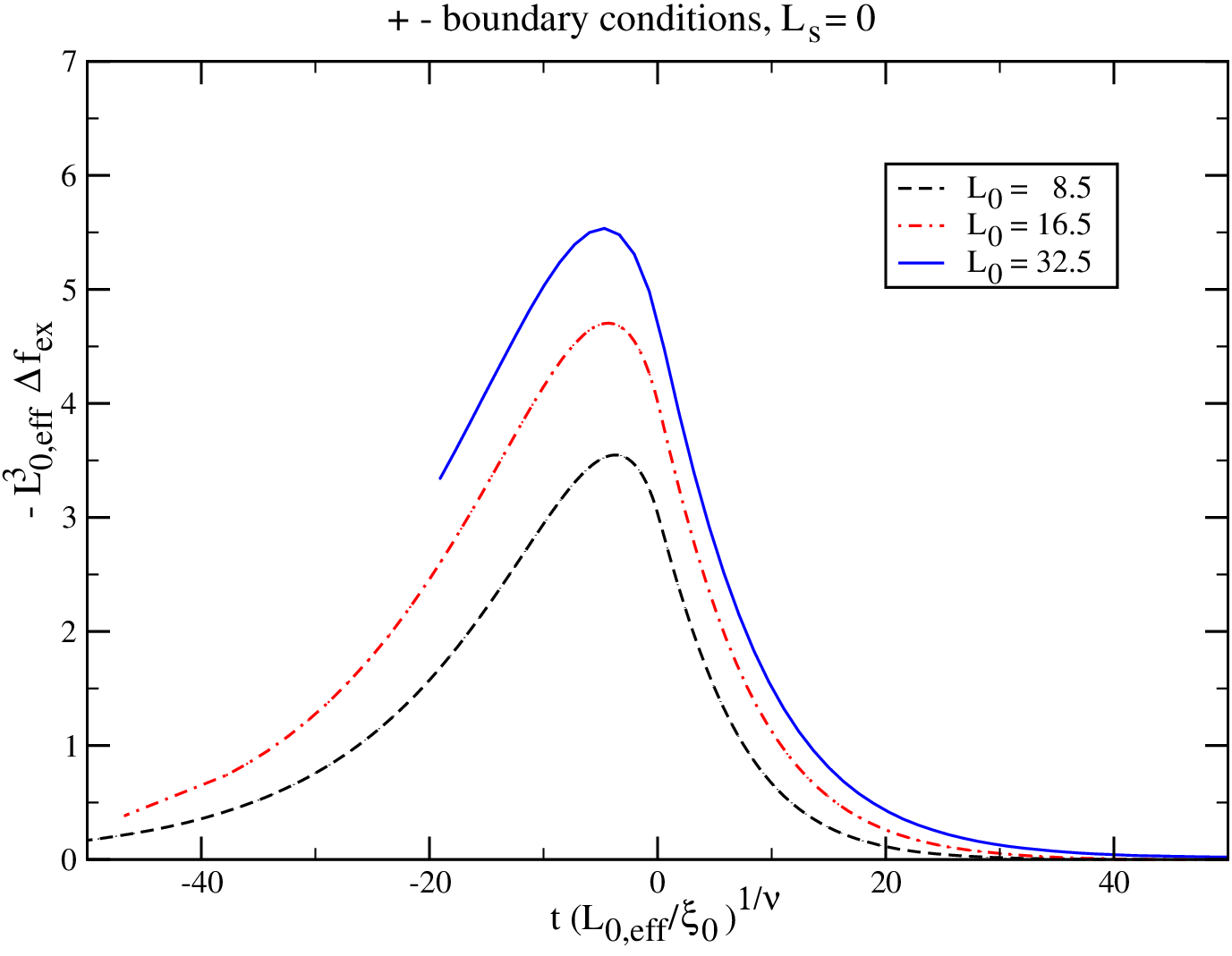}
\vskip1.3cm
\includegraphics[width=13.3cm]{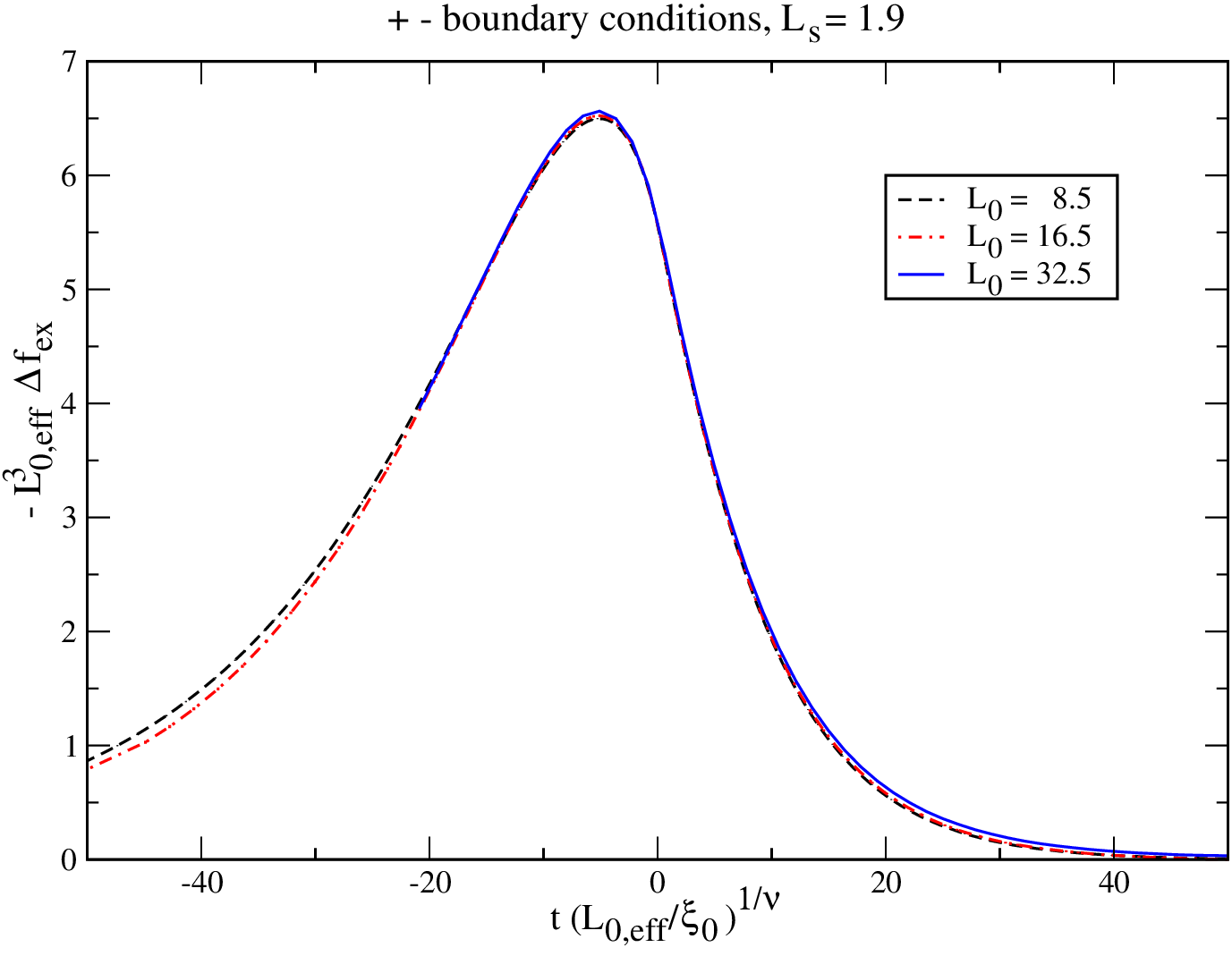}
\caption{\label{antiLs}
$+-$ boundary conditions.
In the upper part of the figure we plot
$- L_{0}^3 \Delta f_{ex}$
as a function of $ t (L_0/\xi_0)^{1/\nu}$
for $L_0=8.5$,  $16.5$ and $32.5$, where we use $\nu=0.63002$ and
$\xi_0=0.2282$. In the lower part we have replaced $L_0$ by 
$L_{0,eff} = L_0+L_s$  with $L_s=1.9$. 
For a discussion see the text.
}
\end{center}
\end{figure}

\begin{figure}
\begin{center}
\includegraphics[width=13.3cm]{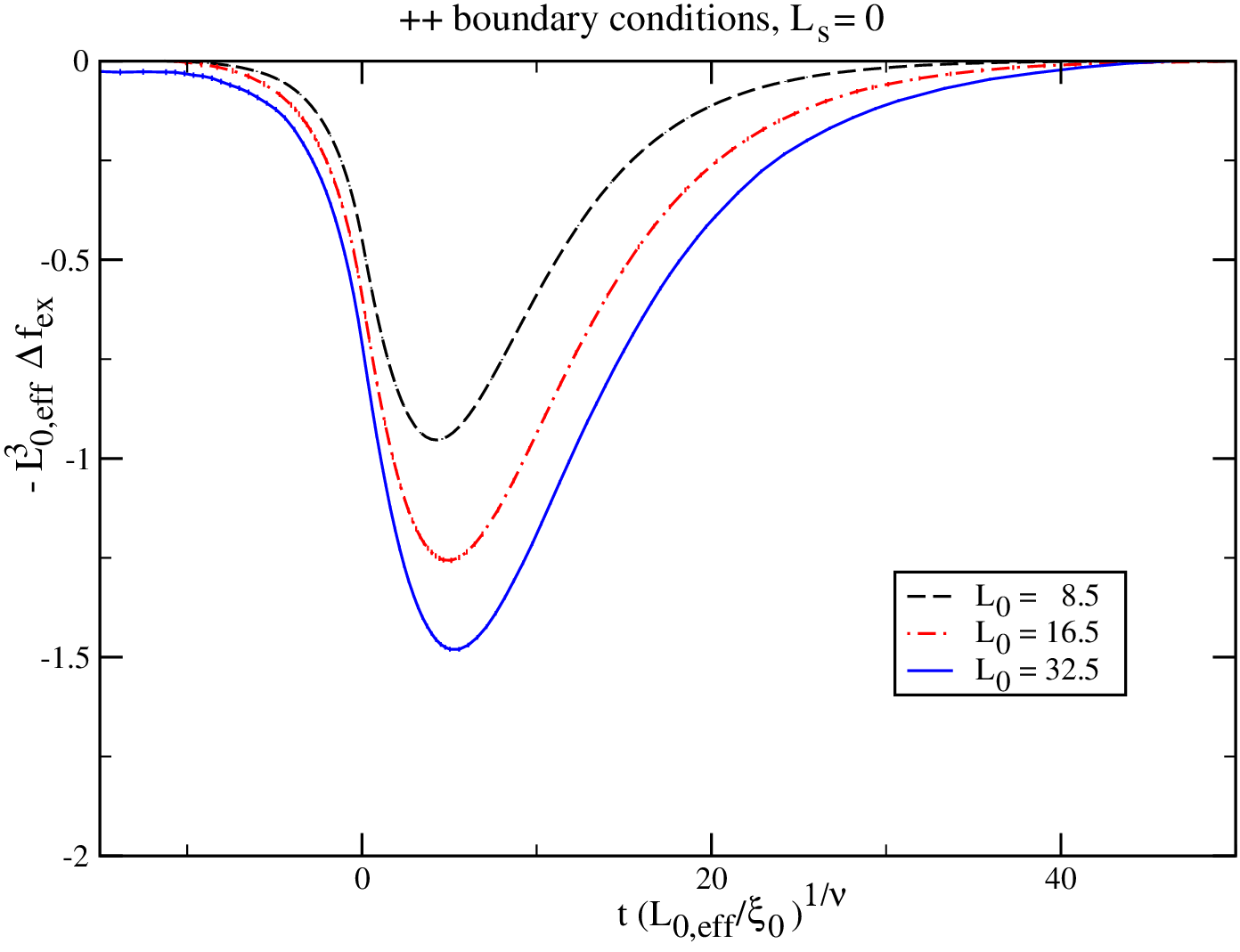}
\vskip1.3cm
\includegraphics[width=13.3cm]{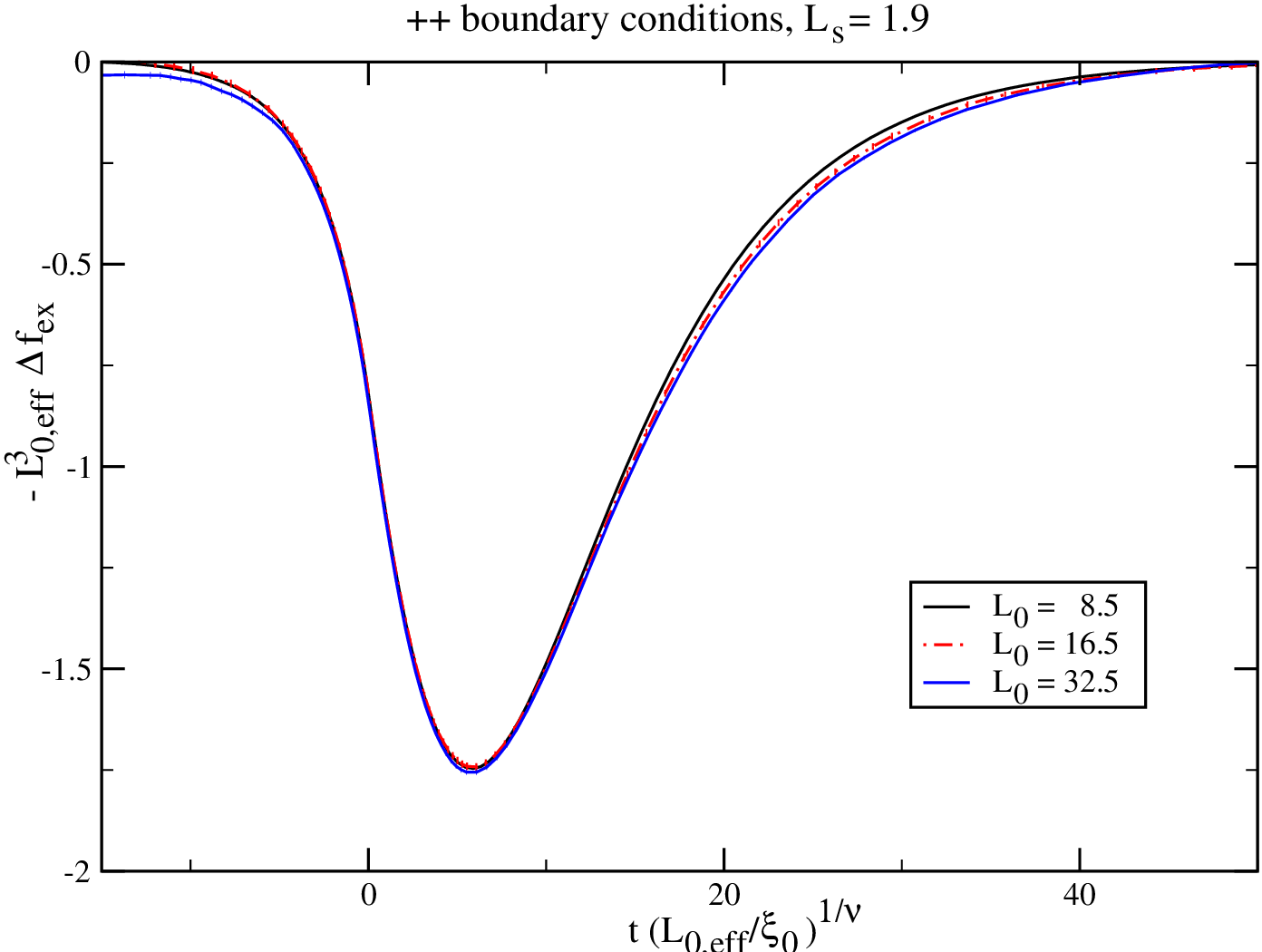}
\caption{\label{paraLs}
$++$ boundary conditions.
In the upper part of the figure we plot
$- L_{0}^3 \Delta f_{ex}$
as a function of $ t (L_0/\xi_0)^{1/\nu}$
for $L_0=8.5$,  $16.5$ and $32.5$, where we use $\nu=0.63002$ and
$\xi_0=0.2282$. In the lower part we have replaced $L_0$ by
$L_{0,eff} = L_0+L_s$  with $L_s=1.9$.
For a discussion see the text.
}
\end{center}
\end{figure}

In figures \ref{antiLs} and \ref{paraLs} we have plotted our results for 
the finite size scaling functions $\theta_{+-}(x)$ and $\theta_{++}(x)$,
respectively. The solid lines
that are plotted linearly interpolate between the data points that we
have computed. Note that
the statistical error of $\Delta f_{ex}(L_0) L_0^3$  is of similar size
as the thickness of the line.
In both cases, in the upper figure we do not take into account
any correction to scaling. This means we  plot $-\Delta f_{ex}(L_0) L_0^3$ as 
a function of $t [L_0/\xi_0]^{1/\nu}$, using $\nu=0.63002$. 

Not taking into account any correction, we see for both $++$ and $+-$ 
boundary conditions a clear discrepancy between the curves for 
$L_0=8.5$, $16.5$ and $32.5$.

Therefore  in the lower part of the figures \ref{antiLs} and \ref{paraLs} 
we have replaced $L_0$ by $L_{0,eff} = L_0 + L_s$, using the value 
$L_s=1.9$ obtained above from the finite size scaling study at the 
bulk critical point. This means that we have plotted 
$-\Delta f_{ex}(L_0) (L_0+L_s)^3$  as a function of 
$t [(L_0+L_s)/\xi_0]^{1/\nu}$.   Now the curves essentially fall on 
top of each other. 
Therefore we do not consider further corrections and 
take the curves obtained for $L_0=16.5$ and $32.5$ as our final result.
The remaining small difference between $L_0=16.5$ and $32.5$ gives us some 
measure for the systematical error of our final result.

Now let us discuss the properties of $\theta_{++}(x)$ and $\theta_{+-}(x)$.
We see that $\theta_{++}(x)$ is  negative and $\theta_{+-}(x)$ is positive 
in the whole range of $x$. This means that in the case of $++$ boundary 
conditions the force is attractive, while for  $+-$ boundary 
conditions it is repulsive.
In both cases the function shows a single extremum. In the case of $++$
boundary conditions it is located in the high temperature phase, while
for $+-$ it is in the low temperature phase. In order to accurately 
locate these extrema, we have computed the zeros of $\Delta E(L_0,\beta)$.
For $++$ boundary conditions we find $\beta_{min}=0.37407(3)$, $0.38219(2)$
and $0.38569(2)$ for $L_0=8.5$, $16.5$ and $32.5$, respectively.  
For these values of $\beta_{min}$ we have computed 
$x_{min}=t_{min} [(L_0+L_s)/\xi_0]^{1/\nu}$ and correspondingly 
$\theta_{min} =- \Delta f_{ex}(\beta_{min}) (L_0+L_s)^3$. As our final 
result we take the value obtained for $L_0=32.5$ using $L_s=1.9$, $\nu=0.63002$ 
and $\xi_0=0.2282$. We arrive at 
\begin{equation}
 x_{++,min} =  5.82(10)   \;\;\; \;\; \theta_{++,min} = - 1.76(3)  \;\;,
\end{equation}
where the quoted error takes into
account the statistical error and the errors due to the uncertainties of
$L_s$, $\xi_0$ and $\nu$.
 
For $+-$ boundary conditions we find $\beta_{max} = 0.39961(2)$, $0.39256(2)$ 
and $0.389525(10)$ for $L_0 =8.5$, $16.5$ and $32.5$, respectively. In the 
same way as above for $++$ boundary conditions we arrive at
\begin{equation}
 x_{+-,max} = - 5.17(7)   \;\;\; \;\; \theta_{+-,max} = 6.56(10)  \;\;.
\end{equation}
At the bulk critical point we get $\theta_{++}(0)=0.84(2)$ and 
$\theta_{+-}(0)=5.56(7)$. These results are less precise but fully 
consistent with those obtained in the previous section, 
eqs.~(\ref{thetapp},\ref{thetapm}).

In ref. \cite{myZZZZZ} we have demonstrated at the example of films 
with periodic and free boundary conditions in the three dimensional XY 
universality class that the relation $\theta(x) = 2 h(x) - \frac{x}{\nu} h'(x)$,
 eq.~(\ref{important}), can be employed to compute  $\theta(x)$ from the 
excess energy per area of the film, without taking the derivative with
respect to the thickness $L_0$ of the film.

The main practical problem of this approach is that for free 
boundary conditions as well as symmetry breaking boundary conditions that 
are studied here, the analytic part of the free energy per area and 
hence also of the energy per area suffers from a boundary correction that
is not described by $L_{0,eff}=L_0+L_s$ of the singular part.
In section \ref{eneatbc} we have already determined the value of 
this correction at the 
bulk critical point. However it turns out that it is not sufficient here
to approximate this correction by a constant. Even by adding a term 
linear in the reduced temperature $t$ to the analytic boundary correction, 
we could not reliably compute $\theta_{++}(x)$ and $\theta_{+-}(x)$. 
We made no attempt to improve this by adding higher order terms. 

\subsection{Behavior at large $|x|$}
\begin{figure}
\begin{center}
\includegraphics[width=13.3cm]{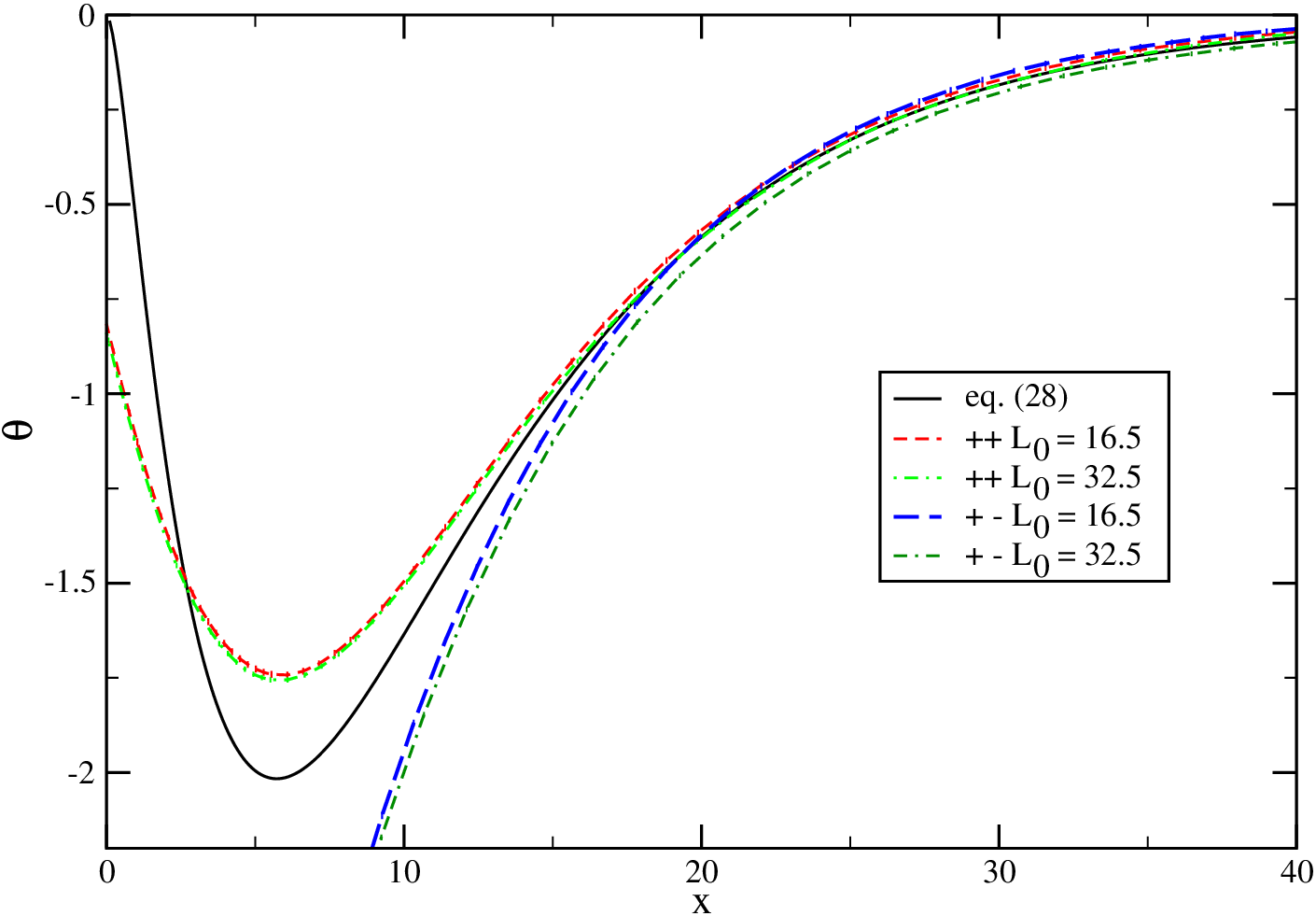}
\caption{\label{comphigh} 
We plot our numerical results for $\theta_{++}(x)$ and $-\theta_{+-}(x)$ 
obtained with $L_0=16.5$ and $32.5$ using $L_s=1.9$ for $x>0$. 
For comparison we give 
the result of eq.~(\ref{scalinghigh}), setting $C^2=1.5$. 
For a discussion see the text.
}
\end{center}
\end{figure}

In figure \ref{comphigh} 
we have plotted $\theta_{++}(x)$ and $-\theta_{+-}(x)$  in the 
high temperature phase. For comparison we have plotted $\theta_{++}(x)$ 
given by eq.~(\ref{scalinghigh}). We have fixed the constant $C^2$ by 
matching the value at $x\approx 20$, where $\theta_{++}(x)$ and 
$-\theta_{+-}(x)$ still agree within the error bars.  We find 
\begin{equation}
\label{C2result}
 C^2 = 1.5(1) \;\;.
\end{equation}

Indeed for $x \gtrapprox 20 $ at the level of our accuracy 
$\theta_{++}(x)$ and $-\theta_{+-}(x)$ are equal. In the same range, the 
two curves are well approximated by eq.~(\ref{scalinghigh}). 

\begin{figure}
\begin{center}
\includegraphics[width=13.3cm]{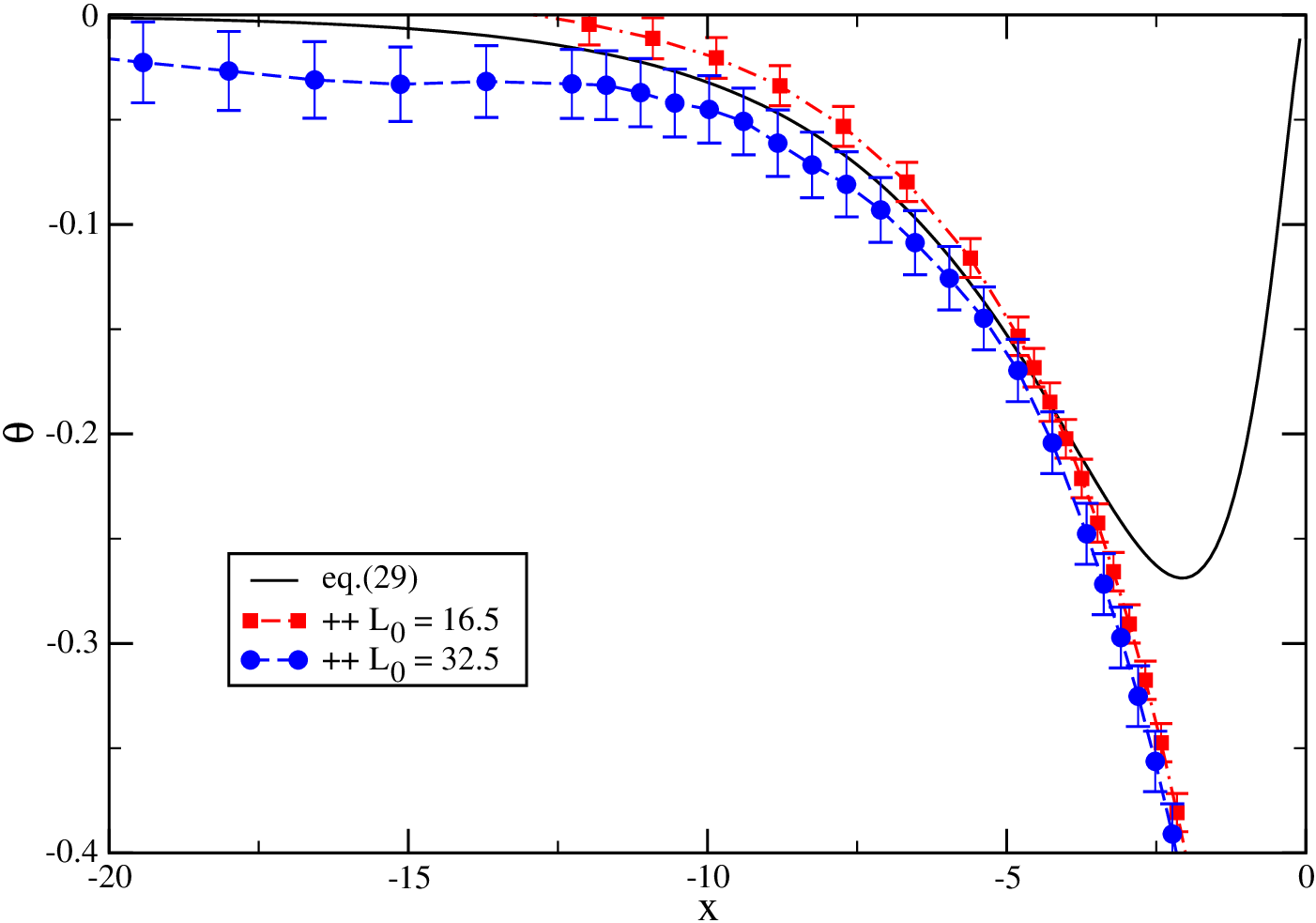}
\caption{\label{complow}
We plot our numerical results for $\theta_{++}(x)$ 
obtained with $L_0=16.5$ and $32.5$ using $L_s=1.9$ for $x<0$. 
For comparison we give
the result of eq.~(\ref{scalinglow}), setting $\bar{C}^2=0.2$.
For a discussion see the text.
}
\end{center}
\end{figure}

Next let us turn to the low temperature phase.  We have matched 
eq.~(\ref{scalinglow}) with our numerical results obtained for $L_0=16.5$ and
$32.5$ and $++$ boundary conditions at $x \approx -7$. We get
\begin{equation}
\bar{C}^2 = 0.20(5)  \;\;.
\end{equation}
As one can see from figure \ref{complow} there is reasonable match between 
our numerical results for $\theta_{++}(x)$ and eq.~(\ref{scalinglow})
for $x \lessapprox  -5$. 
In figure \ref{complow} we have plotted the statistical error of our results. 
The fact that for small $x$, within less than two standard deviations, the 
estimate of $\theta_{++}(x)$ computed for $L_0=16.5$ and $L_0=32.5$ becomes
equal to zero is a non-trivial validation of our numerical integration.

\subsection{Correlation length of the films}
For all simulations discussed above we have measured the 
second moment correlation length as defined in section \ref{xisection}. 
The correlation length is interesting for practical purpose, since 
we have to choose the lattice size $L$ in 1 and 2 direction such that
$L \gg \xi_{2nd}$ in order to avoid sizable effectively two dimensional
finite size effects.
Furthermore we shall discuss the finite size scaling behavior of the 
second moment correlation length of the film to further probe the theoretical
expectations on corrections to scaling.

\begin{figure}
\begin{center}
\includegraphics[width=13.3cm]{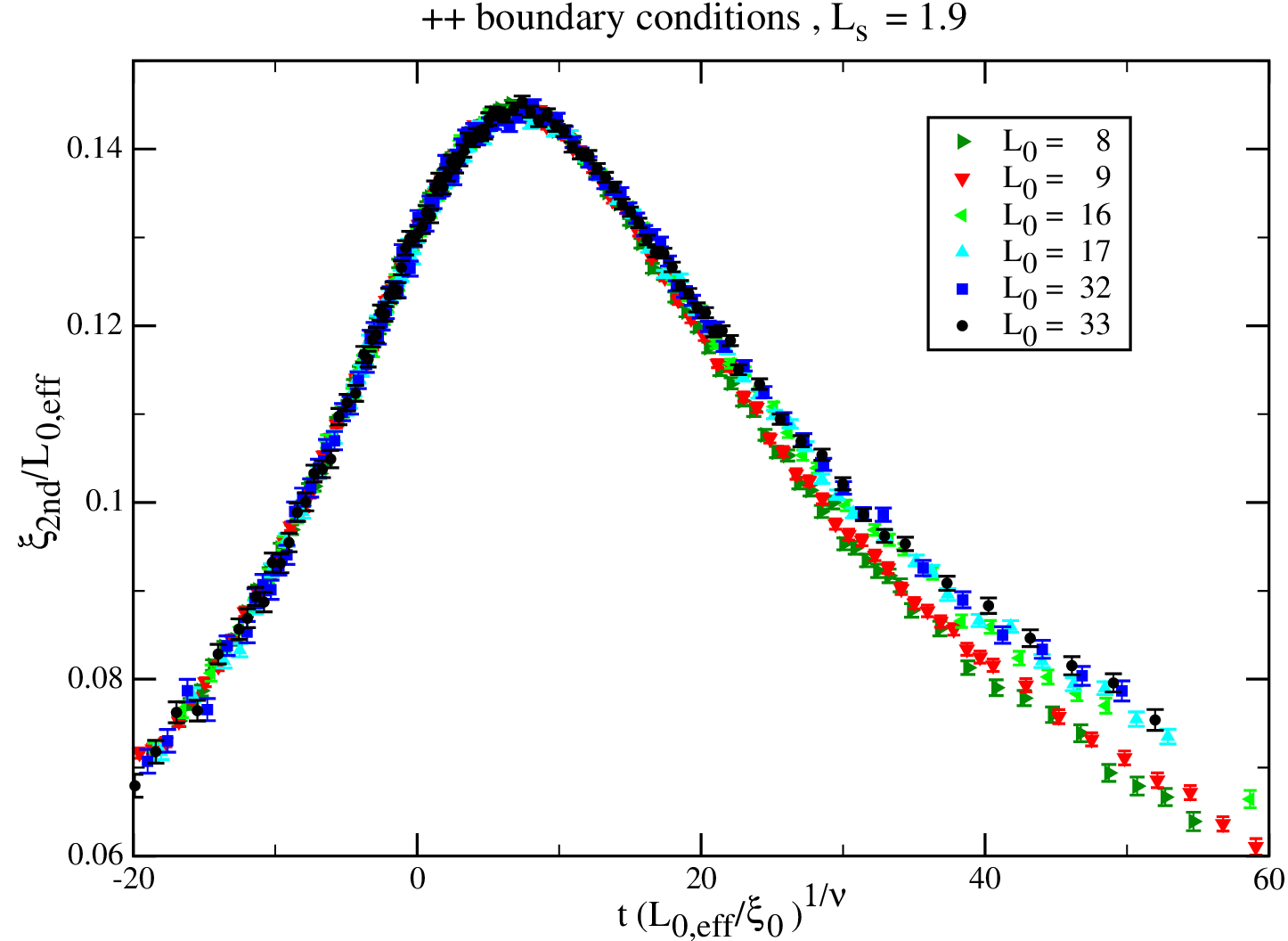}
\caption{\label{xppplot}
For $++$ boundary conditions, we plot $\xi_{2nd}/L_{0,eff}$ as a function
of the scaling variable $x=t [L_{0,eff}/\xi_0]^{1/\nu}$ for $L_0=8$, $9$,
$16$, $17$, $32$ and $33$ using $L_s=1.9$.  Notice that $\xi_{2nd}$ is the 
second moment correlation length of the film, while $\xi_0$ appearing in the 
scaling variable $x$ is the amplitude of the correlation length of the 
bulk system in the high temperature phase.
}
\end{center}
\end{figure}

To this end, we have plotted in figure \ref{xppplot} for $++$ boundary 
conditions  $\xi_{2nd}/L_{0,eff}$ of the film as a function of the 
scaling variable  $x=t [L_{0,eff}/\xi_0]^{1/\nu}$ for the thicknesses 
$L_0=8$, $9$, $16$, $17$, $32$ and $33$.  Using $L_s=1.9$ instead of 
$L_s=0$ clearly improves the collapse of the curves obtained from different 
thicknesses $L_0$.  Using $L_s=1.9$, in the range 
$-20 \gtrapprox x \gtrapprox 20$ the curves obtained for different thicknesses
fall on top of each other within the error bars. For larger values of 
$x$ there is some discrepancy between the thicknesses $L_0=8$ and $9$ and 
$L_0=16$, $17$, $32$ and $33$ on the other hand. This can be attributed to 
analytic corrections to scaling. For all thicknesses  $\xi_{2nd}/L_{0,eff}$
assumes a single  maximum at $x \approx 7$. 

\begin{figure}
\begin{center}
\includegraphics[width=13.3cm]{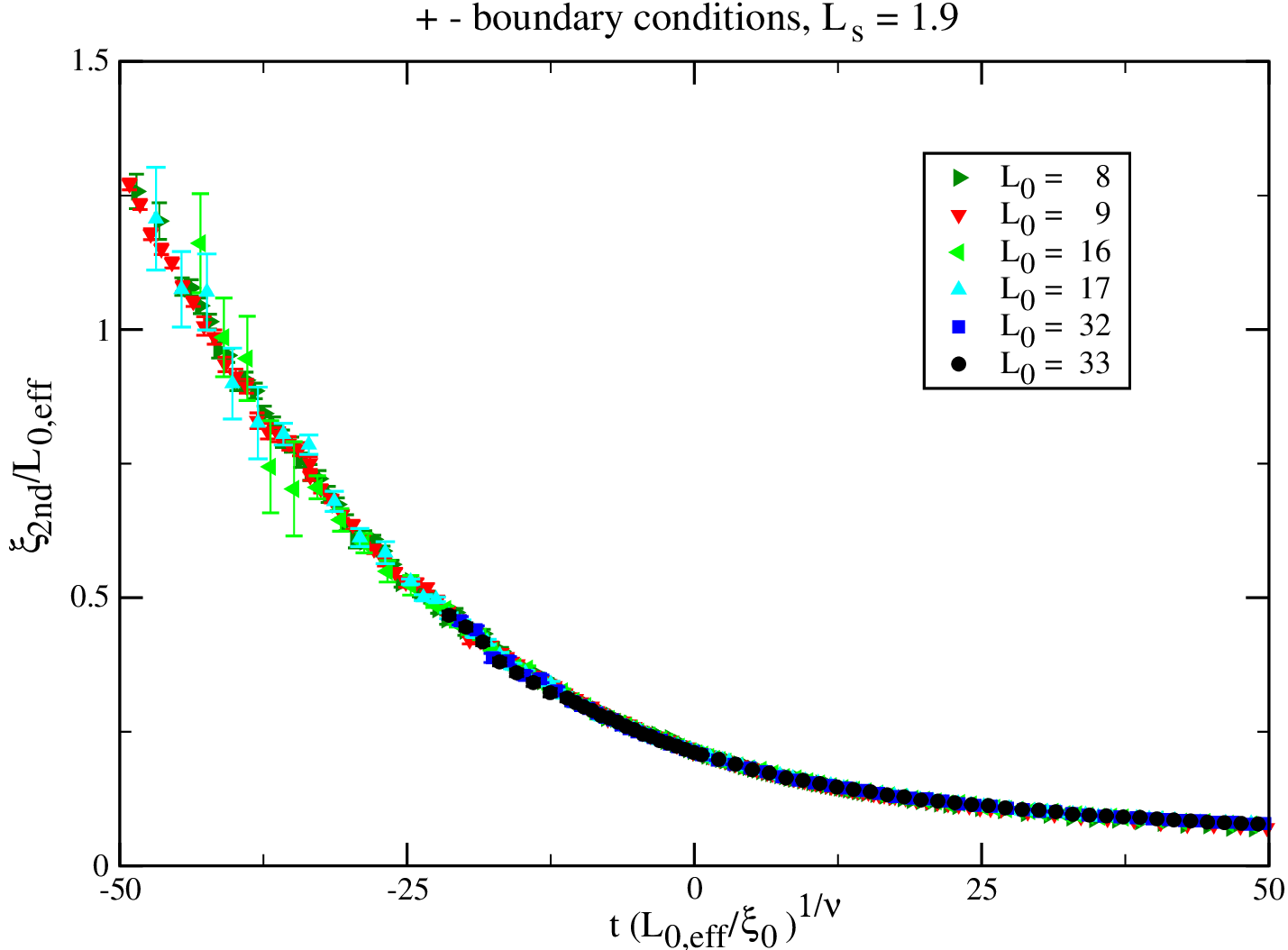}
\caption{\label{xpmplot}
Same as figure \ref{xppplot} for $+-$ instead of $++$ boundary conditions.
}
\end{center}
\end{figure}

Figure \ref{xpmplot} is the analogue of figure \ref{xppplot}  for 
$+-$ instead of $++$  boundary conditions. Also here we find, using 
$L_s=1.9$ a nice collapse  of the curves obtained for the different 
thicknesses of the films. Now  $\xi_{2nd}/L_{0,eff}$ is monotonically
increasing with decreasing $x$. In figure  \ref{xpmplot}  we have 
stopped, a bit arbitrary,  at $x=-50$. For  $x \approx -79.7$, the smallest
value of $x$ that we have reached for $L_0=9$,  we get 
$\xi_{2nd}/L_{0,eff} \approx 3.5$.

With an increasing correlation length the autocorrelation time of the 
Metropolis update increases. Therefore simulations become increasingly difficult as 
we go deeper into the low temperature phase, towards smaller values of 
$x$. 
As a consequence  we had to stop at 
$x \approx -20.4$ and $-21.3$ for $L_0=32$ and $33$, respectively.

\section{Comparison with other theoretical results and experiments}
The scaling functions $\theta_{++}$ and $\theta_{+-}$ have been computed 
recently by using Monte Carlo simulations of the spin-1/2 Ising model
on the simple cubic lattice \cite{VaGaMaDi07,VaGaMaDi08}.  The results are 
presented in figures 3 and 4 of  \cite{VaGaMaDi07} and 
9 and 10 of  \cite{VaGaMaDi08} for $++$ and $+-$ boundary conditions,
respectively. For both types of boundary conditions, 
the final result depends strongly on the 
precise form of the ansatz, see eqs.~(18,20,21,23) of \cite{VaGaMaDi08}, 
for corrections to scaling that is chosen.
Qualitatively, the curves for both $++$ and $+-$ boundary conditions agree 
with ours. 
For the position of the extrema the authors of \cite{VaGaMaDi08} quote
$x_{++,min}=5.90(8)$ and $x_{+-,max}=-5.4(1)$ in the caption of their 
figures 9 and 10, respectively.  
These are in quite good agreement with our results. In 
\cite{Nature,GaMaHeNeHeBe09}, see the discussion below 
eq.~(14) of \cite{GaMaHeNeHeBe09},
the authors extract the amplitude $C^2$ from the data of \cite{VaGaMaDi08}.
Their result depends on the ansatz that is chosen for the corrections and 
also on the boundary conditions. Using the ansatz that is denoted by (i) 
in figures  9 and 10 of \cite{VaGaMaDi08}, they find $C^2=1.51(2)$ and 
$1.82(2)$ for $++$ and $+-$ boundary conditions, respectively. 
Instead, using the ansatz that is denoted by (ii) they arrive at 
$C^2=1.16(2)$ and $1.38(2)$, respectively. It is clear from these numbers
that systematical errors due to corrections to scaling are much larger than 
statistical errors. Taking this into account, there is nice agreement 
with our estimate $C^2 = 1.5(1)$, eq.~(\ref{C2result}).

\begin{figure}
\begin{center}
\includegraphics[width=13.3cm]{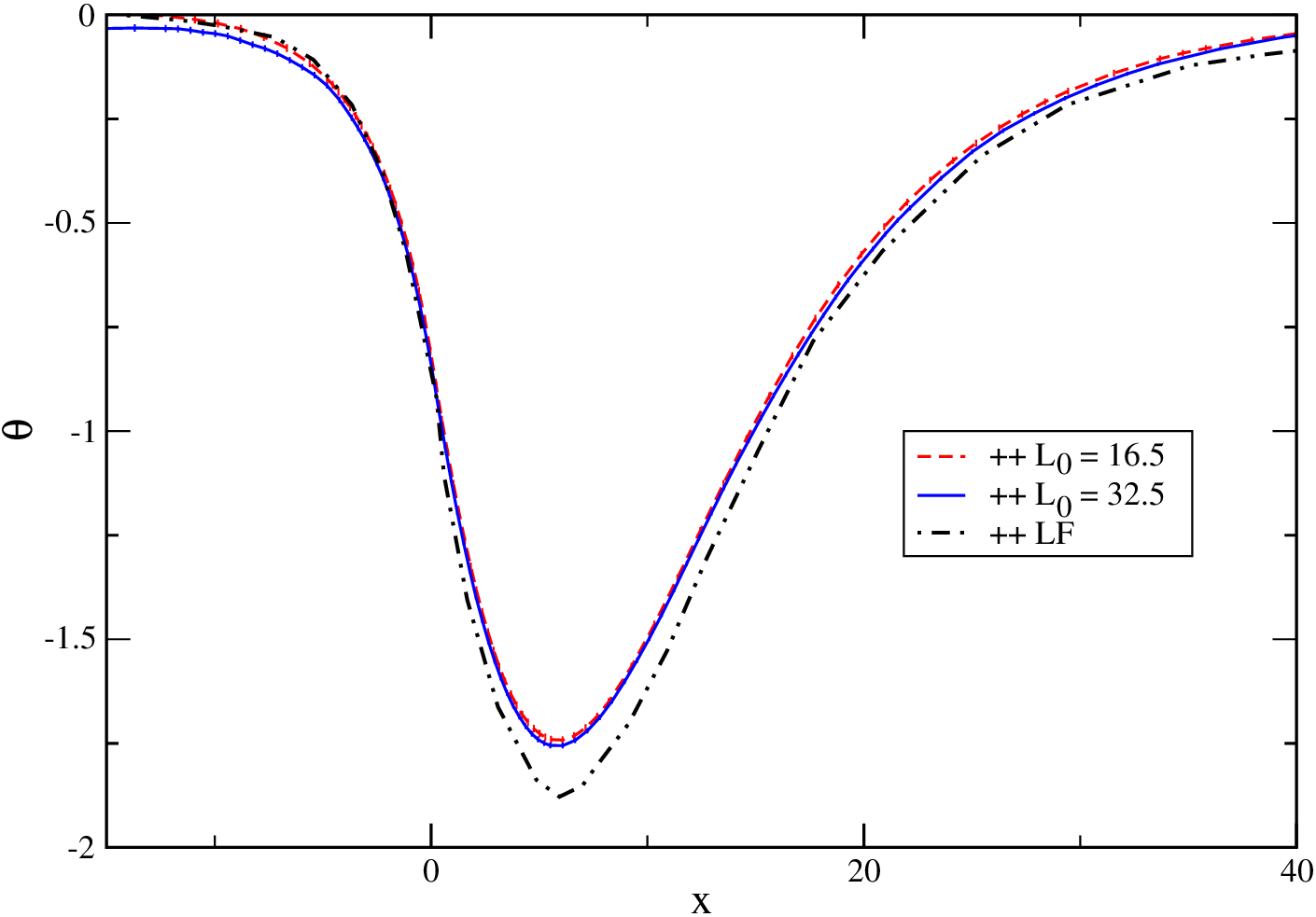}
\caption{\label{upton}
We plot the result of ref. \cite{BoUp08} for $\theta_{++}(x)$ obtained
by using the de Gennes-Fisher local-functional (LF)  method. We have 
copied the curve from fig. 1 of \cite{BoUp08}. 
For comparison we plot our numerical results for $\theta_{++}(x)$
obtained with $L_0=16.5$ and $32.5$ using $L_s=1.9$.
}
\end{center}
\end{figure}

In figure \ref{upton}
we compare our result for $\theta_{++}(x)$ with that obtained 
by using the de Gennes-Fisher local-functional method  \cite{BoUp08}. 
As input the method
uses universal amplitude ratios of the bulk system. Here we made no effort
to redo the calculations of  \cite{BoUp08} using our updated values for 
the  universal amplitude ratios \cite{myamplitude} and value for 
the exponent $\nu$ (Ref. \cite{mycritical}). Instead, we have copied the curve 
from figure 1 of  \cite{BoUp08}. 
Overall we find a reasonable agreement with our result. We see a very small
shift of the  local-functional method curve towards larger values of $x$
compared with ours. Clearly, the value of the minimum  
of the curve obtained by the local-functional method is smaller than that of 
ours.

The authors of \cite{FuYaPe05} have studied wetting films of a 
binary mixture of methylcyclohexane and perfluoromethylcyclohexane.
They have deduced the thermodynamic Casimir force from measurements of the 
thickness of the 
film. Their result for $\theta_{+-}(x)$ given in figure 3 of \cite{FuYaPe05}
is more or less consistent with but much less precise than our result.  
The authors of \cite{Nature,GaMaHeNeHeBe09}  have studied the thermodynamic 
Casimir force between colloidal particles that are immersed into a mixture 
of water and lutidine and the surface of the cell. The surface of the particle
was prepared such that it either preferentially absorbs water or lutidine.
Hence both $++$ and $+-$ boundary conditions were accessible. A major problem 
in the interpretation of the experimental data is to disentangle the 
thermodynamic Casimir
force from other forces.  It turns out that only for relatively large $x$, 
reliable results could be obtained.
Theoretically the colloidal particle and the surface of the cell
are described by a sphere and a plane. In \cite{Nature,GaMaHeNeHeBe09}
the Derjaguin approximation had been used to obtain a prediction for 
this geometry starting from the theoretical results for the universal 
finite size scaling functions $\theta_{++}(x)$ and $\theta_{+-}(x)$ 
for the film geometry. 
The authors of \cite{Nature,GaMaHeNeHeBe09} have fitted their data with the 
equivalent of ansatz~(\ref{scalinghigh}), taking $\xi_0$ as free 
parameter. Their result for $\xi_0$ is consistent with that obtained from 
the analysis of bulk quantities.  This check could be made more 
stringent by replacing the theoretical estimate of $C^2$ of 
\cite{Nature,GaMaHeNeHeBe09} by ours eq.~(\ref{C2result}).

Finally in table \ref{Camplitude} we have summarized results obtained 
for the scaling functions at the bulk critical point. In the literature, 
results obtained by field theoretic methods \cite{Krech97}, 
the de Gennes-Fisher local-functional  method \cite{BoUp98}, Monte Carlo
simulations \cite{Krech97,VaGaMaDi07,VaGaMaDi08} and experiment \cite{FuYaPe05}
can be found. Mostly, in the original work, the so called Casimir 
amplitude $\Delta=\theta(0)/2$ is quoted. We see that field theoretic methods,
in particular the $\epsilon$-expansion, are not able to provide quantitatively
satisfying results. Those of the  de Gennes-Fisher local-functional  method 
\cite{BoUp98}
are in much better agreement with ours.  The results of previous Monte 
Carlo simulations differ by more than the quoted error bars from our
results. Note that in ref. \cite{VaGaMaDi07} only the statistical error
is quoted. The numbers quoted for \cite{VaGaMaDi08} are obtained by using 
an ansatz different from that of \cite{VaGaMaDi07}, which explains the 
difference between them. 
In figure 8 of \cite{VaGaMaDi08} the authors give in addition to the results
obtained with their preferred ansatz those obtained by using two alternative 
ans\"atze. From this comparison one might conclude that the systematical error
is larger than the statistical one that we quote in table \ref{Camplitude}. 

As we have
seen here, for the thicknesses that can be studied today, corrections to 
scaling, in particular those caused by the boundaries, are numerically
important. In order to get an accurate result for the scaling limit, 
these corrections have to be properly taken into account.  In the generic 
case, when corrections $\propto L_0^{-\omega}$, with $\omega=0.832(6)$,
and $\propto L_0^{-1}$ are present this is a difficult task.

\begin{table}
\caption{\sl \label{Camplitude}  Comparison of our results for $\theta_{++}(0)$
and $\theta_{+-}(0)$ with those given in the literature. For a discussion 
see the text.
}
\begin{center}
\begin{tabular}{ccll}
\hline
  Ref.           & Method  &  $\theta_{++}(0)$   &  $\theta_{+-}(0)$ \\
\hline
\cite{Krech97}  & $\epsilon$-expansion&  -0.346    &    3.16          \\
\cite{Krech97}  & $d=3$ expansion &  -0.652    &    4.78          \\
\cite{BoUp98}   & local-functional&  -0.84(16)            &    6.2  \\  
\cite{FuYaPe05} & experiment &     --           &    6(2)          \\
\cite{Krech97}  & Monte Carlo  &  -0.690(32)   &    4.900(64)     \\
\cite{VaGaMaDi07}& Monte Carlo &  -0.884(16)    &       5.97(2)      \\ %%L=13
\cite{VaGaMaDi08}& Monte Carlo &  -0.75(6)     &    5.42(4)        \\
  here           & Monte Carlo &  -0.820(15)   &    5.613(20)      \\
\hline
\end{tabular}
\end{center}
\end{table}

\section{Summary and Conclusions}
We have studied the thermodynamic Casimir force for thin films in the 
three dimensional Ising universality class. In particular 
we have studied  symmetry  breaking boundary conditions. We consider
the two cases $++$ and $+-$, where the fixed spins at the boundary are 
either all positive or are positive at one boundary and negative at the 
other.
We have simulated the improved Blume-Capel model on the simple cubic 
lattice.  The boundary conditions are expected to cause corrections 
that are to leading order $\propto L_0^{-1}$. In general it is 
hard to disentangle such corrections from leading corrections to finite size 
scaling which are $\propto L_0^{-\omega}$ where $\omega=0.832(6)$ 
(Ref. \cite{mycritical}). In the improved model, corrections to scaling 
$\propto L_0^{-\omega}$ are eliminated. This fact very much simplifies the
analysis of the Monte Carlo data. In particular we could clearly  demonstrate 
that the corrections caused by the boundaries can be expressed
by an effective thickness  $L_{0,eff}= L_0+L_s$. For our model we find, 
for both $++$ and $+-$ boundary conditions $L_s = 1.9(1)$.

Having corrections to scaling well under control, we have obtained
reliable results for the universal finite size scaling functions 
$\theta_{++}(x)$ and $\theta_{+-}(x)$,
where $x=t [L_{0,eff}/\xi_0]^{1/\nu}$, of the thermodynamic Casimir force.
For large values of $x$, 
we have compared our estimates for $\theta_{++}(x)$ and $\theta_{+-}(x)$
with the prediction~(\ref{scalinghigh}) derived by using the transfer matrix
formalism. We find good agreement. For large values of $-x$ we have compared 
$\theta_{++}(x)$ with eq.~(\ref{scalinglow}) also  derived by using the 
transfer matrix formalism. Also here we find agreement. 

Finally we have compared our estimates for $\theta_{++}(x)$ and $\theta_{+-}(x)$
with field theoretic calculations, the de Gennes-Fisher local-field method,
previous Monte Carlo simulations and experiments. While field theory does
not provide quantitatively satisfying results, those of the local-field method
are in quite reasonable agreement with ours. Also
the results of previous Monte Carlo simulations are essentially in agreement
with ours. 

\section{Acknowledgements}
This work was supported by the DFG under the grant No HA 3150/2-1.

\end{document}